\documentclass[aps,pra,10pt,twocolumn,floatfix,nofootinbib,superscriptaddress,longbibliography,nobibnotes]{revtex4-2}

\usepackage{amsmath}
\usepackage{mathtools}
\usepackage{wasysym}
\usepackage{newtxmath}

\usepackage{array} 
\usepackage[T1]{fontenc}
\usepackage{newtxtext}
\usepackage{dsfont}                 
\usepackage{bbold}                  
\usepackage[normalem]{ulem}         

\usepackage{enumerate}
\usepackage[shortlabels]{enumitem}

\usepackage[dvipsnames,table]{xcolor}
\usepackage{graphicx}
\graphicspath{{figs/}}

\usepackage[flushleft]{threeparttable}
\usepackage{multirow}

\usepackage{physics}                
\usepackage{csquotes}               
\usepackage{comment}

\usepackage[colorlinks=true,citecolor=cyan,linkcolor=magenta,filecolor=magenta]{hyperref}
\usepackage[capitalize]{cleveref}

\usepackage{tikz}
\usetikzlibrary{decorations.pathmorphing}
\usetikzlibrary{arrows,decorations.markings,cd}

\usepackage[caption=false]{subfig}
\usepackage{orcidlink}

\DeclareMathAlphabet{\mathbbold}{U}{bbold}{m}{n}

\newcommand{\be}{\begin{equation}}
\newcommand{\ee}{\end{equation}}

	\def\ket#1{|{#1}\rangle}		

\usepackage{bm}

\definecolor{TB}{rgb}{0.93,0.47,0.2}
\def\TB#1{{\color{TB}#1}}

\begin{document}


\title{Superconductivity in hyperbolic spaces:\texorpdfstring{\vspace{1mm}\\}{} Regular hyperbolic lattices and Ginzburg-Landau theory}

\author{Vladimir Bashmakov}
\email{vladimir.bashmakov@ru.nl}
\affiliation{Institute for Molecules and Materials, Radboud University, Heyendaalseweg 135, 6525AJ Nijmegen, \mbox{The Netherlands}}
\author{Askar Iliasov}
\email{askar.iliasov@uzh.ch}
\affiliation{Department of Physics, University of Zurich, Winterthurerstrasse 190, 8057 Zurich, Switzerland}
\author{Tom\'{a}\v{s} Bzdu\v{s}ek\,\orcidlink{0000-0001-6904-5264}}
\email{tomas.bzdusek@uzh.ch}
\affiliation{Department of Physics, University of Zurich, Winterthurerstrasse 190, 8057 Zurich, Switzerland}
\author{Andrey A. Bagrov\,\orcidlink{0000-0002-5036-5476}}
\email{andrey.bagrov@ru.nl}
\affiliation{Institute for Molecules and Materials, Radboud University, Heyendaalseweg 135, 6525AJ Nijmegen, \mbox{The Netherlands}}


\begin{abstract}
We study $s$-wave superconductivity in hyperbolic spaces using the Bogoliubov-de Gennes theory for discrete hyperbolic lattices and the Ginzburg-Landau theory for the continuous hyperbolic plane. 
Hyperbolic lattices maintain a finite fraction of boundary sites regardless of system size, thus fundamentally altering superconductivity through enhanced boundary effects absent in flat space.
Within the BCS framework for hyperbolic lattices, uniform systems reproduce standard bulk behavior, whereas finite systems with open boundaries, studied through exact diagonalization and Cayley-tree approximations, exhibit boundary-enhanced superconductivity and boundary-only superconducting states that persist above the bulk critical temperature.
Numerical studies further reveal that boundary termination critically determines superconducting properties; in particular, rough boundaries with dangling bonds generate zero-energy modes that raise critical temperatures by several times relative to smooth boundaries.
Turning to the complementary Ginzburg-Landau analysis of the hyperbolic plane, we find that finite geometries permit radial variations of the condensate absent in infinite space.
Owing to the interplay between coherence length and curvature radius, the theory exhibits two types of superconductivity even without magnetic fields, with vortices replaced by lines of vanishing order parameter in the nontrivial type.
Our findings establish hyperbolic geometry as a platform for engineering boundary-controlled superconductivity, opening new directions for physics in curved spaces in condensed matter and holography.

\end{abstract}

\maketitle


\section{Introduction}

The quantum physics of hyperbolic lattices represents a very new and rapidly developing direction in condensed matter physics research. 
Even understanding single-particle quantum mechanics of such structures turned out to be a complex endeavor that resulted in a number of interesting results in a short span of time. 
A major focus has been put on the spectral properties, hyperbolic crystallography, and developing band structure theory of hyperbolic lattices \cite{Boettcher:2022,Boettcher:2022b,Maciejko:2022,Lenggenhager:2023,Rayan:2022, Nagy:2024}, which, in turn, has been applied to studying topological models \cite{Chen:2023,Liu:2022,Tao:2023,Bzdusek:2022,Urwyler:2022,Sun:2024}. 
The spectral properties were studied by different methods as well, such as the Selberg trace formula \cite{Adil:2022}, continued fraction expansion of Green's functions \cite{Mosseri:2023}, and converging periodic boundary conditions \cite{Lux:2022,Lux:2023}.
By considering the magnetic field effect on the band structure and density of states, the dependence of the Hofstadter butterflies shape on the Schl\"{a}fli symbols $\{p,q\}$ \cite{Stegmaier:2022} was analyzed, and the emergence of Dirac cones was demonstrated \cite{Ikeda:2021}. Another activity, where the surface has been barely scratched, is holography of hyperbolic lattices relating single-particle Green's functions to the dual quantum field theory on the edge, which helped establish correspondence between quantum mechanics of hyperbolic lattices and one-dimensional interacting aperiodic systems \cite{Erdmenger:2023,Erdmenger:2023b,Flicker:2020,Asaduzzaman:2020}. There are works on Anderson localization \cite{Chen:2024} and connection to Yang-Mills theories as well \cite{Shankar:2024}. One should also mention that the studies of single-particle physics on hyperbolic lattices are not constrained by theoretical research, but there is a number of exciting electric circuit experiments \cite{Kollar:2019,Lenggenhager:2022,Boettcher:2020,Dey:2024}.

At the same time, study of many-body effects on hyperbolic lattices has been limited to very few results. In particular, interacting quantum spin models have been considered \cite{Gotz:2024, Lenggenhager:2024gmf,Dusel:2025,Mosseri:2025,Vidal:2025vlz}. 
In Ref.~\citenum{Gotz:2024}, emergence of global antiferromagnetic order on hyperbolic lattices has been studied, and, among other things, it was shown that despite the fact that finite-depth hyperbolic patches are dominated by effectively one-dimensional boundaries, quantum fluctuations do not destroy order even at the boundary. 
In Refs.~\citenum{Lenggenhager:2024gmf,Dusel:2025,Mosseri:2025,Vidal:2025vlz}, Kitaev model on hyperbolic lattices has been introduced and solved exactly, and the existence of different types of spin liquid phases in its ground state phase diagram was shown. Itinerant electrons have also been analyzed. 
In Ref.~\citenum{Roy:2025}, it was shown that, depending on the Schl\"{a}fli indices, hyperbolic lattices can host Dirac liquids, Fermi liquids, or lead to emergence of flat bands. Metal-insulator transition driven by finite-size effects has been analyzed there as well.

\begin{figure}
 \centering
    \includegraphics[width=\linewidth]{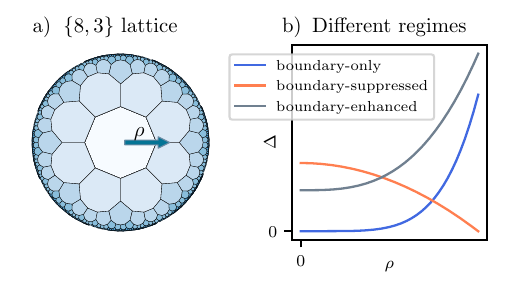} 
    \caption{(a) An example of $\{8,3\}$ hyperbolic lattice is shown in the panel. (b) Possible behavior of the order parameter $\Delta$ with increasing the distance from the center $\rho$ is shown in the panel.}
    \label{fig:hyper_sketch}
\end{figure}

With this paper, we continue the endeavor of studying many-body physics on hyperbolic lattices, and put focus on superconductivity motivated by the following two aspects of hyperbolic geometry. First, hyperbolic lattices are an example of expander graphs, with any finite-depth hyperbolic patch having an excessive number of boundary sites comparable or exceeding that of bulk sites. This makes them a perfect playground to study edge physics and speak about boundary superconductivity in the spirit of Ref.~\citenum{Babaev:2020} and consequent works \cite{Croitoru:2020,Hainzl:2022,Barkman:2022,Talkachov:2023}, where the formation of boundary superconducting states with enhanced and suppressed critical temperatures was shown for conventional BCS superconductors.
Secondly, crystallography of hyperbolic space goes far beyond that of the Euclidean plane. While in zero-curvature space only square, triangular, and honeycomb lattices are allowed in two dimensions (if we speak of tiling the plane with identical polygons), the hyperbolic plane can be tiled by regular polygons with any number of edges, as long as the basic condition $(p-2)(q-2)>4$ is satisfied (here $p$ is the number of edges 
of the polygons, and $q$ is the site degree). 
Such regular tesselation of a hyperbolic plane is called a regular hyperbolic lattice and can be characterized by the
Schläfli symbol $\{p,q\}$. The example case of $\{8,3\}$ lattice is illustrated in Fig.~\ref{fig:hyper_sketch}(a).
Hyperbolic lattices provide a freedom to study, in a more systematic way, the role of geometric frustrations and rather smoothly interpolate between Euclidean lattice geometries. 
Another possibility is the appearance of unconventional order parameters due to non-Euclidean hyperbolic point groups~\cite{Chen:2023} and non-Abelian nature of the translation group of the hyperbolic lattices~\cite{Boettcher:2022}. 

In our study, we adopt two-fold approach and consider both discrete lattices (Sec.~\ref{sec:lattices}), which we analyze within the Bogoliubov-de Gennes (BdG) framework by focusing on the attractive Hubbard model, and the continuous hyperbolic plane (Sec.~\ref{sec:ginzburg-landau}), which we investigate through the lens of the Ginzburg-Landau theory in a curved space. 
Both approaches lead to consistent results, revealing the critical role of boundaries, untypical for the flat space. First, as summarized by Fig.~\ref{fig:hyper_sketch}(b), we find that, depending on the boundary conditions, superconducting order parameter can acquire diverse spatial distributions corresponding to boundary-only superconductivity, boundary-enhanced superconductivity, and boundary-suppressed superconductivity.  
In the discrete regime, we show that appropriate choice of boundary geometry can lead to emergence of a large number of zero modes, which, in turn, drastically enhances superconductivity leading to a several times higher critical temperature and condensate density. We also scan the parameter space of the Hubbard model at different $\{p,q\}$, showing how departing from the flat-space discretization ($\{6,3\}$, $\{4,4\}$, $\{3,6\}$) alters its phase diagram. 

A paper complementary to this addresses related topics such as mean-field analysis of $s$-wave superconductivity on Cayley trees and the Bardeen-Cooper-Schrieffer (BCS) theory in continuous hyperbolic space~\cite{Pavliuk:2025}.

\section{Hyperbolic Lattices}\label{sec:lattices}

\subsection{Uniform lattices}\label{sec:uni_lat}

The starting point of our study of the $s$-wave superconductivity is the Hubbard model with on-site attractive interactions
\begin{equation}\label{eq:Hubbard_model}
    H=t\sum_{\langle i,j \rangle\sigma} c^{\dagger}_{i\sigma}c_{j\sigma}-\mu\sum_{i\sigma}n_{i\sigma}-U\sum_{i} n_{i\uparrow} n_{i\downarrow}
\end{equation}
where $c^\dagger_{i\sigma},\,c_{i\sigma}$ are creation and annihilation operators of a fermion with spin $\sigma=\uparrow,\downarrow$ at site $i$, $n_{i\sigma}=c^{\dagger}_{i\sigma} c_{i\sigma}$ is the particle number operator, $t=-1$ is the nearest-neighbor hopping amplitude, $\mu$ is the chemical potential, and $U>0$ is the strength of an attractive on-site interaction.

We investigate the Hubbard model on the mean-field level by employing the (BdG) approach \cite{BdG_book}. 
After performing Bogolubov transformation and introducing local pairing amplitude $\Delta_i$, the eigenvalue equation for the BdG Hamiltonian $\mathcal{H}$ reads
\begin{gather}
    \mathcal{H}\begin{pmatrix}
    u_{n}\\v_{n}
    \end{pmatrix}=\begin{pmatrix}
    h-\mu&\delta\\\delta&-h+\mu
    \end{pmatrix}    \begin{pmatrix}
    u_{n}\\v_{n}
    \end{pmatrix}=E_{n}\begin{pmatrix}
    u_{n}\\v_{n}
    \end{pmatrix},
\label{eq:BdG_Ham}
\end{gather}
where $h$ is the matrix representing the free part of the Hamiltonian (i.e., the adjacency matrix of a lattice multiplied by $t$), $\delta$ is a diagonal matrix consisting of values of the order parameter $\delta=\mbox{diag}(\Delta_1,\dots\Delta_N)$ (the diagonal structure is a consequence of the on-site Hubbard interaction), and $N$ is the total number of lattice sites. 
We assume the absence of spin-orbit coupling and $s$-wave symmetry of the order parameter, leading us to omit the spin indices. 
The corresponding self-consistent equations for the superconducting order parameter $\Delta$ are as follows~\cite{BdG_book} 
\begin{gather}
    \Delta_i=\frac{U}{2}\sum_{n}u_{n,i} v^{*}_{n,_i} \tanh(\frac{E_n}{2 T}), \label{eq:selfcons_delta}
\end{gather}
where the sum is over all eigenstates of $\mathcal{H}$.

 \begin{figure}
 \centering
    \includegraphics[width=\linewidth]{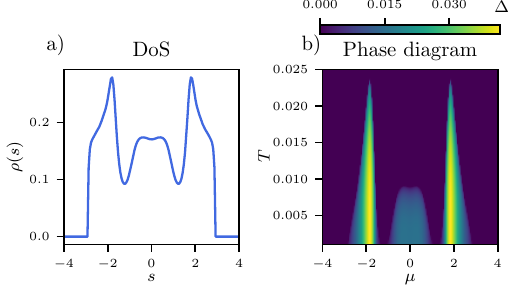} 
    \caption{(a)~Density of states and (b)~the phase diagram for the $\{8,3 \}$ hyperbolic lattice. The Hubbard potential is~$U=1$.}
    \label{fig:8_3_example}
\end{figure}

Before considering the more general case of hyperbolic lattices with open boundary conditions, let us comment on the simpler case of site-transitive graphs. These correspond to lattices (including hyperbolic $\{p,q\}$ lattices with periodic boundary conditions) that are uniform in the sense that any two of their sites are related by a symmetry. In such a case, assuming that the crystalline symmetry is not broken, the order parameter $\Delta_i \equiv \Delta$ is uniform~\cite{Pavliuk:2025}, and the self-consistent equation can be brought to the closed integral form
\be\label{eq:uniform_gap}
\!\Delta=U\!\!\int^{\infty}_{-\infty}\!\frac{\Delta\rho(\lambda)}{2\sqrt{(\lambda-\mu)^2+\Delta^2}}\tanh\!\Big(\frac{\sqrt{(\lambda-\mu)^2+\Delta^2}}{2 T}\Big)d\lambda,
\ee
where $\rho(\lambda)$ is the (uniform) density of states on the lattice. 
For a lattice with a boundary, the site transitivity is a justifiable approximation in the bulk but it explicitly fails near the boundary; hence, these calculations explicitly capture bulk superconducting properties. 
The obtained gap equation~(\ref{eq:uniform_gap}) reproduces BCS theory, implying that
the critical temperature depends on the interaction potential as $T_{c}\sim e^{-\frac{1}{U\rho(\mu)}}$.

We can use the gap equation given by Eq.~\eqref{eq:uniform_gap} to calculate the superconducting gap for an infinite hyperbolic lattice. 
In our computations, we approximate the bulk density of states for the hyperbolic lattice using the continued fraction expansion of the Green's function~\cite{Mosseri:2023}.
Results for the density of states of the $\{8,3 \}$ hyperbolic lattice and the corresponding phase diagram are shown in Fig.~\ref{fig:8_3_example}. 

Since the bulk of both Euclidean and hyperbolic lattices are uniform in the specified sense, the resulting description of superconductivity in the bulk is qualitatively equivalent for both geometries. As a next step, one might wonder if the equivalence still applies in the presence of a boundary. In this regard, although the boundary breaks the uniformity in both cases, one should note that these boundaries comprise a drastically different extent of the lattice. In particular, the boundary of a Euclidean lattice becomes negligibly small in the thermodynamic limit. In contrast, hyperbolic lattices are expander graphs~\cite{Placke:2025}, implying that their boundary constitutes a finite fraction of the volume irrespective of the system size. This motivates us to investigate the behavior of the superconducting order near the boundary of a hyperbolic lattice. 

This section continues further with three other subsections. In Sec.~\ref{sec:latt-small}, we consider small-size hyperbolic lattices amenable to exact diagonalization. In Sec.~\ref{sec:Cayley_tree_apx}, we find a formal way to approximate certain hyperbolic lattices with Cayley trees of fractional site degree, which allows us to study systems of a large depth. In Sec.~\ref{sec:BdG}, we numerically solve the BdG equations for the attractive Hubbard model at arbitrary chemical potential and diverse $\{p,q\}$.

\subsection{Hyperbolic lattices with small radial size}
\label{sec:latt-small}
 \begin{figure}
 \centering
    \includegraphics[width=\linewidth]{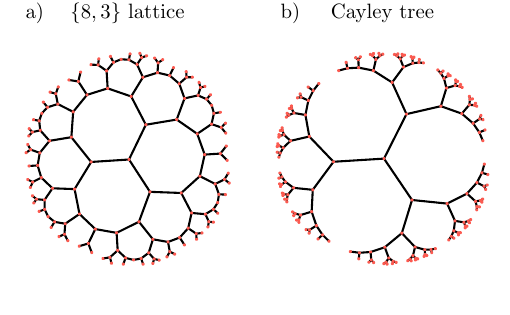}
    \vspace{-0.8cm}
    \caption{Illustrations of (a) site-centered flake of $\{8,3\}$ hyperbolic lattice and of (b) Cayley tree with site degree $q=3$ (connectivity $K=2$). Both lattices are shown in the Poincaré disc representation. }
    \label{fig:hyper_tree_examples}
\end{figure}

In this section, we consider the results of exact diagonalization of site-centered hyperbolic lattices and compare them against Cayley trees, i.e., rooted tree graphs with a uniform site degree (except for sites at the boundary, whose site degree is one). We contrast the tree-like structure of the Cayley tree with site degree $q=3$ against the hyperbolic $\{8,3\}$ lattice with the same site degree in Fig.~\ref{fig:hyper_tree_examples}. 

To proceed, we thus construct a hyperbolic flake starting from a central site and keep all sites whose graph distances (counted in the number of nearest-neighbor edges) are less than or equal to some specified bound. 
We compare superconductivity in such hyperbolic flakes against Cayley trees with matching site degree and the same graph radius. Throughout this section, we consider the particular choice of the chemical potential $\mu=0$ (i.e., half-filling) unless specified otherwise.

The solution of the self-consistent BdG equations on Cayley trees is fully radially symmetric. 
The introduction of cycles in hyperbolic lattices breaks the exact radial symmetry; nonetheless, one may expect that the radial symmetry holds approximately. 
To characterize the degree to which the radial symmetry persists
on hyperbolic lattices, we consider the variation of the radial order parameter, defined by the formula
\be
\label{eqn:Delta-var}
\sigma_l=\sqrt{\left(\sum_{i \in S_l}\frac{1}{N_l}\Delta_i\right)^2-\frac{1}{N_l}\sum_{i\in S_l} \Delta^2_i},
\ee
where $S_l$ denotes the $l$-th shell of the lattice (i.e., the collection of sites with graph distances $l$ from the center) and $N_l$ is the number of sites within this shell.

\begin{figure}
    \centering
    \includegraphics[width=\linewidth]{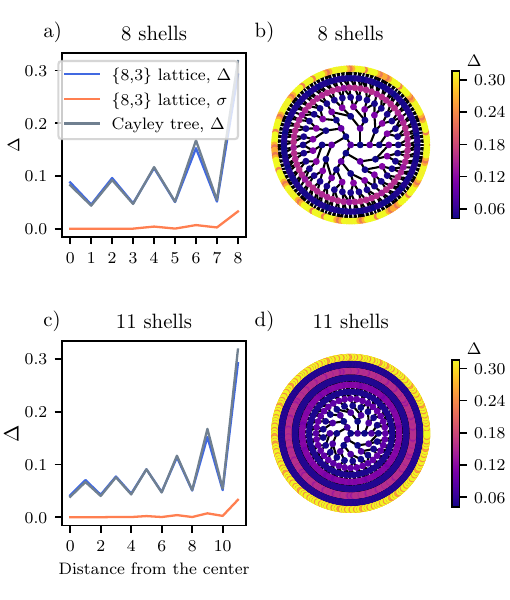}
    \caption{Spatial distribution of the superconducting order parameter $\Delta$ on a finite $\{8,3\}$ hyperbolic lattice with $8$ resp.~$11$ shells for parameters $T=0.01$, $\mu=0$ and $U=1$. 
    To obtain the radial profile of $\Delta$, the order parameter is averaged over shells. In panels (a) and (c), the radial dependence of $\Delta$ is also shown for the Cayley trees with the same site degree $K+1=3$ and with the same number of shells.}     \label{fig:8_3_lattices_Delta}
\end{figure}

First, we compare the order parameters calculated for lattices with $8$ shells. 
We consider the $\{8,3\}$ hyperbolic lattice and the Cayley tree with connectivity $K=2$ (connectivity of a tree is equal to the degree of the bulk sites minus one: $K=q-1$). 
Results of the calculations for parameters $T=0.01$, $\mu=0$, and $U=1$ are shown in Fig.~\ref{fig:8_3_lattices_Delta}. 
Let us note that, in contrast to the Poincaré disc model used in Fig.~\ref{fig:hyper_tree_examples}, we adopt in Fig.~\ref{fig:8_3_lattices_Delta}(b,d) a different representation of the hyperbolic lattice. Namely, we embed the lattice into a plane such that all sites in shell $S_l$ are placed on a disc with radius $l$ around the center (while making sure that adjacent sites remain connected by an edge). In this way, the boundary effects are more easily accessible from a visual perspective. We will call this the `radial embedding' of the lattice.

Figure~\ref{fig:8_3_lattices_Delta} demonstrates that the profile of the averaged order parameter on the hyperbolic lattice is similar to the profile of the Cayley tree, and the values of the order parameter are close to each other. 
One should further observe that the variation of the order parameter within a shell of the hyperbolic lattice is relatively small, and slightly increases only closer to the boundary.
These features are observed for lattices with both $M=8$ [Fig.~\ref{fig:8_3_lattices_Delta}(a,b)] and $M=11$ [Fig.~\ref{fig:8_3_lattices_Delta}(c,d)] shells. 
Finally, let us point out that the boundary variation and the boundary value of the order parameter for the larger hyperbolic lattice are close to the values calculated on the smaller lattice; in contrast, the order parameters near the center is noticeably smaller for the lattice with the larger radius.

\begin{figure}
    \centering
    \includegraphics[width=\linewidth]{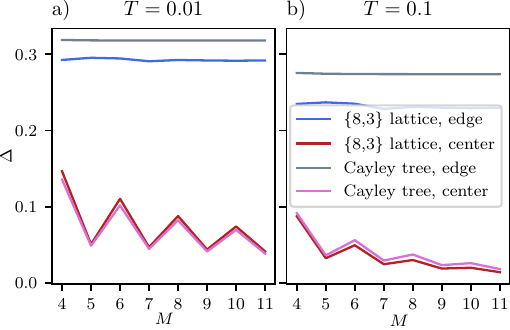}
  
    \caption{The dependence of the superconducting order parameter $\Delta$ on the radial size $M$ of the lattice. 
    The value of $\Delta$ at the edge and at the center of the $\{8,3\}$ hyperbolic lattice as well as of a Cayley tree with $K+1=3$ are shown. 
    For the hyperbolic lattice, the order parameter at the edge is averaged over the shell. The displayed data is obtained for parameters $\mu=0$ and $U=1$, and the temperature is set to (a)~$T=0.01$, resp.~to (b)~$T=0.1$.}
    \label{fig:various_M}
\end{figure}

 \begin{figure}
    \centering
    \includegraphics[width=\linewidth]{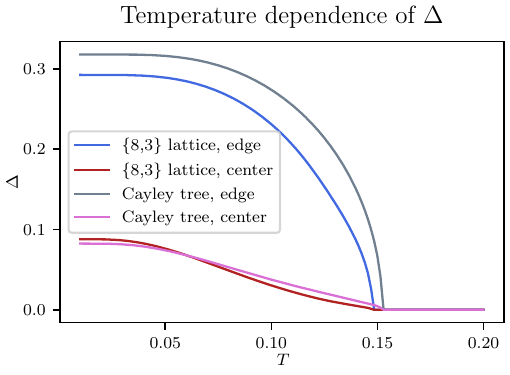}
  
    \caption{ The temperature dependence of $\Delta$ (for specified paramaters) for a flake of the $\{8,3\}$ hyperbolic lattice and a Cayley tree with $K+1=3$. Both lattices have $8$ shells, the parameters $\mu=0$ and $U=1$. The order parameters at the center of the lattice and at the last shell are plotted. For the hyperbolic lattice, the order parameter at the last shell is averaged over the shell.}
    \label{fig:finite_hyplat_phase}
\end{figure}

We proceed with a more detailed comparison studying various radial sizes. 
The dependence of the superconducting order parameter for the $\{8,3\}$ hyperbolic lattice and for the Cayley tree with connectivity $K=2$ on the total number of layers $M$ is shown in Fig.~\ref{fig:various_M}. 
Specifically, we plot the values of the superconducting parameter at the center and at the edge for parameters $U=1$ and $\mu=0$, and at two distinct temperatures $T=0.01$ [Fig.~\ref{fig:various_M}(a)] and $T=0.1$ [Fig.~\ref{fig:various_M}(b)].
In the case of the hyperbolic lattice, the order parameter at the edge is averaged over the outermost shell. We observe that the order parameters calculated for the hyperbolic lattices and for the Cayley trees exhibit similar behavior for all radial sizes. 
Notably, boundary values of the order parameter appear stable under a change of the lattice size. 
In contrast, the order parameter in the center decreases with increasing radial size. 
This observation corresponds to the fact that, for both temperatures, as radius $M$ grows, the order parameter at the center converges to the result for an infinite lattice. 
These considerations also suggests that the order parameter in the bulk is smaller than the order parameter at the boundary.

Another similarity between the superconducting phases on the two lattices is demonstrated by Fig.~\ref{fig:finite_hyplat_phase}, where the temperature dependence of the order parameter (for fixed $\mu=0$ and $U=1$) is shown
for the $\{8,3\}$ hyperbolic lattice with $8$ shells and for the Cayley tree with connectivity $K=2$ and $8$ shells. The figure compares the order parameters at the central site and at the edge. 
One can see that the temperature dependence of the two lattices follows similar trends at both locations. 
A particular feature of both lattices, at least for the specified system size, is that the critical temperature at the central node appears to be the same as the critical temperature at the boundary for both lattices, and that the order parameter on the boundary is enhanced.

The observed similarities between small hyperbolic lattices and Cayley trees allow us to anticipate that larger hyperbolic lattices should also have a boundary phase distinguished from the bulk phase, in analogy with the same phenomenon predicted for Cayley trees~\cite{Pavliuk:2025}. 
In the following discussion, we leverage the described analogies to develop an efficient Cayley-tree approximation of the superconducting order parameter on hyperbolic lattices. The developed approximation will allow us to consider significantly larger systems, going far beyond the possibilities of an exact diagonalization description. 
With this approach, we will argue that boundary-only superconductivity can emerge on hyperbolic lattices in a certain range of temperatures.

\subsection{Cayley-tree approximation}
\label{sec:Cayley_tree_apx}

The study of hyperbolic lattices with small radius suggests that the variation of the order parameter along the angle coordinate is relatively small, even for larger systems. 
Inspired by this observation, we introduce in this section an effective radially symmetric model for hyperbolic lattices. 
The simplest way to proceed along this line would be to build an effective Cayley tree representing a hyperbolic lattice.
If successful, such an approximation would allow us to consider much larger radial sizes and to better distinguish between the bulk and the boundary behavior. 
The reason thereof is that, since Cayley trees enjoy exact radial symmetry, one can decompose the Hamiltonian into a block-diagonal form by virtue of a symmetry-adapted basis (which consists of so-called symmetric and nonsymmetric basis states) as developed in Refs.~\cite{Mahan-2001-PRB,Aryal:2020,Ostilli:2022,Hamanaka:2024} (see Appendix~\ref{app:Cayley_BdG} for the construction of this basis and details of the algorithm employing the block decomposition). 
Therefore, in order to construct the Cayley-tree approximation of a hyperbolic lattice, one needs to understand how to generalize (non)symmetric states.

To proceed, let us first focus on the symmetric states. 
In the case of the Cayley tree, one can introduce these states by averaging over shells. 
For a general radially symmetric graph, we can consider the symmetric states in the same manner. 
Let us label the shells of a graph as $S_l$, such that $S_0$ corresponds to the central site, and we denote the number of sites in the $l$-th shell as $N_l$. 
Then, we define the non-normalized symmetric states as follows
\be
|l)'=\sum_{\alpha\in S_l} |\alpha\rangle.
\ee
By construction, these states are orthogonal to each other. Now, we impose that the following relations hold
\begin{gather}
\label{eq:H_on_unnorm_states}
H|0)'=|1)'\\\nonumber
H|l)'=|l+1)'+c_l |l-1)',
\end{gather}
where $H$ is a nearest-neighbor Hamiltonian defined on a radially symmetric graph, and $c_l$ are some constants. 
In other words, the action of the Hamiltonian on non-normalized symmetric states transfers the state to the neighboring symmetric states.\footnote{Let us remark
that our Hamiltonian contains no hopping processes that would transfer the state to more distant layers or within the same layer, although such processes are admissible in radially symmetric graphs.}
However, while the stated relations (\ref{eq:H_on_unnorm_states}) hold on Cayley trees, they do not directly apply to hyperbolic lattices, as these lack exact radial symmetry.
Nevertheless, since we have observed that the order parameter on hyperbolic lattices enjoys an approximate radial symmetry [cf.~the plots of $\sigma$ in Fig.~\ref{fig:8_3_lattices_Delta}(a,c)], it seems reasonable to expect the relations~(\ref{eq:H_on_unnorm_states}) to be \emph{approximately} correct for hyperbolic lattices (at least for the model parameters considered thus far).
Thus, as the defining principle of the Cayley-tree approximation of hyperbolic lattices, we impose Eq.~(\ref{eq:H_on_unnorm_states}) to hold.

Since the Hamiltonian $H$ is Hermitian, the relations in Eq.~\eqref{eq:H_on_unnorm_states} are sufficient to obtain the action of the Hamiltonian on normalized symmetric states $|l)$, defined as
\be\label{eq:effsym_state}
|l)=\frac{1}{\sqrt{N_l}}\sum_{\alpha\in S_l} |\alpha\rangle.
\ee
Specifically, by considering the analogy with Cayley trees [cf.~Eq.~(\ref{eq:H_sym})] one obtains that the components of the Hamiltonian block in the sector of symmetric basis states are the fractions between the normalization factors for the symmetric states between two consecutive shells. 
Hence, we can write the effective Hamiltonian for the generalized symmetric states as follows
\begin{gather}\label{eq:H_eff_tree}
    h^\textrm{eff}_\textrm{sym} = \begin{pmatrix}
0 & \sqrt{N_1} & 0 & 0 & \cdots \\
\sqrt{N_1} & 0 & \sqrt{\frac{N_{2}}{N_1}} & 0 & \cdots \\
0 & \sqrt{\frac{N_2}{N_1}} & 0 & \sqrt{\frac{N_3}{N_2}} & \cdots \\
0 & 0 & \sqrt{\frac{N_3}{N_2}} & 0 & \cdots \\
\vdots & \vdots & \vdots & \vdots & \ddots \\
\end{pmatrix},
\end{gather}
where $N_l$ is the number of sites in the shell $l$, and the values $c_l$ in Eq.~\eqref{eq:H_on_unnorm_states} are equal to $c_l=\frac{N_l}{N_{l-1}}$. 
We see that this block exactly corresponds to the Hamiltonian of the symmetric sector on the Cayley tree for which $N_l=(K+1)K^{l-1}$. 
By comparing the Hamiltonian in Eq.~\eqref{eq:H_eff_tree} with the symmetric sector of the Cayley tree, we can also infer the effective connectivity number for the Cayley-tree approximation of radially symmetric graphs. 
Specifically, the effective non-integer connectivity depends on the shell index
and equals $\tilde {K}_l =\frac {N_{l+1}}{N_l}$. 

Before further developing the Cayley-tree approximation, let us note that, for a hyperbolic lattice, the question of finding $N_l$ is related to determining the growth function of the non-abelian translation group of the lattice. The numbers $N_l$ can then be efficiently calculated by using the series expansion of the growth functions~\cite{Wagreich:1982, Cannon:1992, Bartholdi:2002}. 
It is also interesting to note that one can find the asymptotic behavior of $N_l$ when $l\to\infty$ by studying so-called site types \cite{Gouezel:2015,Nagnibeda:2023}.
In the particular case of $q=3$, one can also apply the recursive relations given in Ref.~\citenum{Mosseri:2023}, and obtain the asymptotic behavior of $\widetilde K_l$ for the $\{8,3\}$ lattice: $\lim_{l\to\infty}\widetilde K_l=\frac{1}{4}(1+\sqrt{13}+\sqrt{2\sqrt{13}-2})\simeq 1.722$.

The observation that one can introduce non-integer connectivity $\tilde{K}_l$ at the level of symmetric states allows us to generalize the non-symmetric states as well. 
The non-symmetric states are defined for a given `seed' node $\beta$, and take non-zero values only in some branch of a Cayley tree emanating from the `seed'. 
In particular, the sectors of generalized non-symmetric states should be captured by 
Hamiltonian blocks similar to the non-symmetric blocks of
the Cayley tree, but with non-integer connectivity $\tilde {K}_l$ that depends on the shell index (and with the initial index appearing in the matrix shifted by the distance of the seed node from the center.) 
Therefore, for the Hamiltonian of the generalized non-symmetric sector, we can write
\begin{gather}
    h^\textrm{eff}_{\textrm{non-sym}} = \begin{pmatrix}
0 & \sqrt{\tilde K_j} & 0 & 0 & \cdots \\
\sqrt{\tilde K_j} & 0 & \sqrt{\tilde K_{j+1}} & 0 & \cdots \\
0 & \sqrt{\tilde K_{j+1}} & 0 & \sqrt{\tilde K_{j+2}} & \cdots \\
0 & 0 & \sqrt{\tilde K_{j+2}} & 0 & \cdots \\
\vdots & \vdots & \vdots & \vdots & \ddots \\
\end{pmatrix},
\end{gather}
where $j$ is the distance of the first non-zero shell from the center, which is equal to the distance of `the seed' node $\beta$ from the center plus one: $j=l_\beta+1$.

Finally, we need to adapt the self-consistent equations. 
To that end, we need to account for two different aspects: the multiplier coming from the degeneracy of the non-symmetric states, and the multiplier coming from the unitary rotation from the original position basis to the basis of (non)symmetric states.
We perform these two considerations in two steps.

First, we account for the degeneracy of nonsymmetric states. For usual Cayley trees, the degeneracy of non-symmetric states equals the number of branches emanating from the `seed' minus one. This is the number of possible linearly independent weightings of branches that create destructive interference and sum up to zero at the seed node under the action of the Hamiltonian. 
Let us assume that this relation still works for non-integer connectivity. 
Then the multiplier coming from the degeneracy of generalized non-symmetric states takes the form $\tilde K _ {l_\beta}-1$, where $l_\beta$ is the distance of the `seed' node from the center. 
The resulting approximated self-consistent BdG equations on a hyperbolic lattice read
\begin{gather}
    \label{eq:selfcons_efftree_orbasis}
    \Delta_i=\frac{U}{2}\sum_{\substack{\textrm{symmetric}\\\textrm{states}}}\sum_{n}u_{n,i} v^{*}_{n,i} \tanh(\frac{E_n}{2 T})+\\
    +\frac{U (\tilde K _ {l_\beta}-1)}{2}\sum_{\substack{\textrm{nonsymmetric}\\\textrm{states,}\,\, l_\beta}}\sum_{n}u_{n,i} v^{*}_{n,i} \tanh(\frac{E_n}{2 T}). \nonumber
\end{gather}
Here, we indexed non-symmetric states by the distance of the `seed' node from the center. In the obtained self-consistent equations, the information about the Schläfli symbol $\{p,q\}$ of the hyperbolic lattice has been 
contained in the non-integer connectivity $\tilde K_{l_\beta}$.

 \begin{figure}[t]
    \centering
    \includegraphics[width=\linewidth]{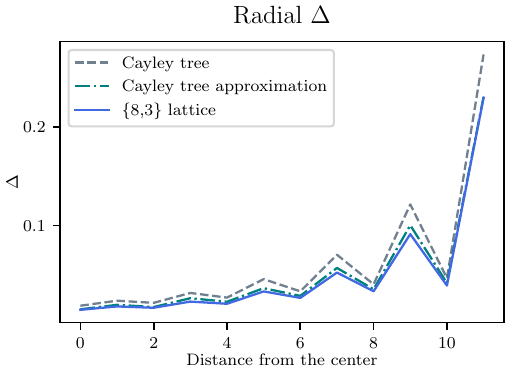} 
    \caption{The spatial profile of the order parameter for the Caylee tree with connectivity $K=2$, for the $\{8,3\}$ hyperbolic lattice (with the order parameter averaged over the shells), and its Cayley-tree approximation. The parameters are $U=1$, $\mu=0$, and $T=0.1$.
    }
    \label{fig:radial_Deltas}
\end{figure}

 \begin{figure}[t]
    \centering
    \includegraphics[width=\linewidth]{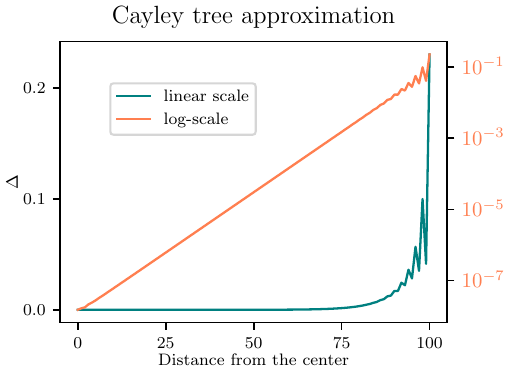} 
    \caption{The spatial profile of the order parameter for the Caylee-tree approximation of the $\{8,3\}$ hyperbolic lattice for the total number of layers $M=100$. The other parameters are kept at $U=1$, $\mu=0$, and $T=0.1$.}
    \label{fig:Delta_Caylee_tree_effective}
\end{figure}

 \begin{figure}[t]
    \centering
    \includegraphics[width=\linewidth]{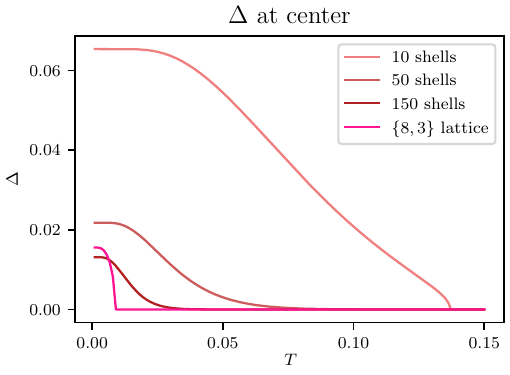} 
    \caption{
    The temperature dependence of $\Delta$ at the center of the Cayley-tree approximations of $\{8,3\}$ lattice for different radial sizes. 
    The figure also compares the results of the approximation against the thermodynamic limit of the infinite $\{8,3\}$ lattice. The parameters are $U=1$ and $\mu=0$.}
    \label{fig:Deltas_at_center}
\end{figure}

 \begin{figure}[t]
    \centering
    \includegraphics[width=\linewidth]{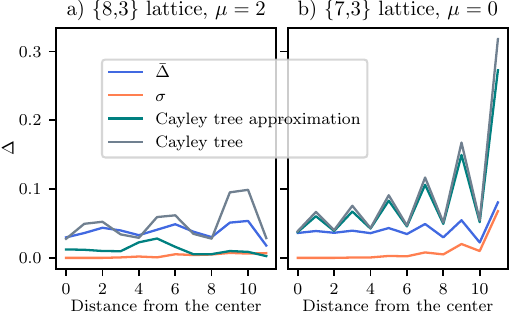}
  
    \caption{The dependence of the order parameter $\Delta$ for different hyperbolic lattices and chemical potentials. Panel (a) shows the averaged $\Delta$ over shells and its variation $\sigma$ for $\{8,3\}$ lattice with radial size $M=11$, and $\mu=2$. The panel also shows the radial dependence of $\Delta$ for the Cayley-tree approximation of $\{8,3\}$ lattice. Panel (b) shows the averaged $\Delta$ over shells and its variation $\sigma$ for $\{7,3\}$ lattice with radial size $M=11$, and $\mu=0$. The panel also shows the radial dependence of $\Delta$ for the Cayley-tree approximation of $\{7,3\}$ lattice. Both panels show the radial dependence of $\Delta$ for the Cayley tree with $K+1=3$ as well. For both panels $U=1$, $T=0.01$.}
    \label{fig:efftree_cases}
\end{figure}

In the next step, we include the effect of the unitary rotation to the symmetry-adapted basis of states. This will allow us to apply an efficient numerical scheme for Cayley trees as outlined in Appendix~\ref{app:Cayley_BdG}. 
The transition from the position basis to the symmetry-adapted basis is given by the averaging as in Eq.~\eqref{eq:effsym_state}. 
Therefore, introducing new BdG vectors $(\widetilde{u}^j, \widetilde{v}^j)$ for $j$-th (non)symmetric sector (where $j$ denotes the index of the first shell in which the state has a non-zero amplitude, i.e., $j=0$ for symmetric states and $l_\beta+1$ for non-symmetric states), we can write the effective self-consistent equations within the Cayley-tree approximations as follows
\begin{gather}
    \label{eq:selfcons_effHL_delta}
    \Delta_i=\frac{U}{2}\sum_{\substack{\textrm{symmetric}\\\textrm{states}}}\sum_{n}\frac{\widetilde u_{n,i} \widetilde v^{*}_{n,i}}{N_i} \tanh(\frac{E_n}{2 T})+\\
    +\frac{U}{2}\sum_{\substack{\textrm{nonsymmetric}\\\textrm{states,}\,\, l_\beta}}\sum_{n}\frac{(N_{l_\beta+1}-N_{l_\beta})\widetilde u_{n,i} \widetilde v^{*}_{n,i}}{N_i} \tanh(\frac{E_n}{2 T}), \nonumber
\end{gather}
where index $i$ denotes the distance from the central node, and index $n$ goes over energy levels for each of the sectors. If the numbers $N_i$ equal the number of shell nodes for some Cayley tree, the self-consistent equations \eqref{eq:selfcons_effHL_delta} reproduce the BdG equations \eqref{eq:selfcons_Cayley} for Cayley trees.
Finally, having implemented the steps above, we can numerically solve Eq.~\eqref{eq:selfcons_effHL_delta} to obtain the profile of the superconducting order parameter on hyperbolic lattices within the Cayley-tree approximation.

To verify the reliability of the developed approximation, we start with showcasing in Fig.~\ref{fig:radial_Deltas} a comparison of the radial order parameter calculated for the Cayley tree with connectivity $K=2$, for the $\{8,3\}$ hyperbolic lattice, and for its Cayley-tree approximation. 
The calculations were made for $U=1$, $\mu=0$, and $T=0.1$. (As usual, we perform averaging over shells for the non-approximated hyperbolic $\{8,3\}$ lattice). 
One can see that the Cayley-tree approximation accurately describes the boundary order parameter of the actual hyperbolic lattice.

We next consider increasing the number of layers $M$ of the (approximated) hyperbolic lattice. 
The profile of the order parameter for the Cayley-tree approximation is shown in Fig.~\ref{fig:Delta_Caylee_tree_effective}, calculated for $U=1$, $\mu=0$, and $T=0.1$. 
The obtained data demonstrates the formation of a superconducting state localized on the boundary: the order parameter decays towards the center of the lattice. This behavior can be understood by comparing with the bulk properties of the infinite system. 
The phase diagram for the infinite $\{8,3\}$ hyperbolic lattice [Fig.~\ref{fig:8_3_example}(b)] indicates that at $U=1$ and $\mu=0$, the bulk critical temperature is approximately $T_c \approx 0.01$. However, the calculation in Fig.~\ref{fig:Delta_Caylee_tree_effective} is performed at $T=0.1$, which is an order of magnitude higher than the bulk $T_c$. At this elevated temperature, the order parameter in the center of a large finite lattice with open boundaries should approach the value corresponding to the infinite lattice, which under these parameters is $\Delta=0$. 
The fact that $\Delta$ remains close to zero near the center in our calculation confirms this expectation, while the finite values at the boundary reflect the boundary superconductivity that persists even above the bulk critical temperature.

The results of more elaborated calculations are shown in Fig.~\ref{fig:Deltas_at_center}. There, we show the values of $\Delta$ at the center of the Cayley-tree approximation of the hyperbolic $\{8,3\}$ lattice for different radial sizes and compare them against the thermodynamic limit of the infinite $\{8,3\}$ lattice. 
While for smaller tree approximations we find the critical temperature to be significantly higher than the result for the infinite lattice, one can observe that for larger tree approximations the values of the central $\Delta$ approach $0$ in an extended range of temperatures.
Interestingly, we also observe that
the values of the central $\Delta$ 
in the bulk-superconducting regime as obtained from the Cayley-tree aproximation are 
\emph{smaller} than the value predicted for
the infinite $\{8,3\}$ lattice.
This observation suggests that the tree approximation is more accurate for the description of superconductivity near the edge than in the bulk \footnote{This can be explained by the fact that the Cayley-tree approximation captures the overall spectrum of the hyperbolic lattice (which is going to be dominated by the boundary) at $\mu=0$ better than it captures the local denisty of states at the center.}. 

Another observation is that in the previous examples, the order parameter exhibits period-$2$ oscillations. 
Owing to a comparison with Cayley trees, where such period-$2$ oscillations can be derived analytically~\cite{Pavliuk:2025}, we anticipate the period-$2$ oscillations observed for the hyperbolic lattice to be a consequence of half-filling in a system with spectrum symmetric around $\mu = 0$.
However, period-$2$ seems incompatible with odd loops as well as with other filing factors (i.e., $\mu\neq 0$ chemical potential). Therefore, one can expect the uniformization approach described in this section to be limited and applicable only for even-$p$ hyperbolic lattices at chemical potential around $\mu=0$. 
This is indeed the case, as demonstrated by Fig.~\ref{fig:efftree_cases}. 
Therein, we show the result for the $\{8,3\}$ lattice and $\mu=2$, Fig.~\ref{fig:efftree_cases}(a) and for the $\{7,3\}$ lattice with $\mu=0$, Fig.~\ref{fig:efftree_cases}(b).
We compare the exact results with their corresponding effective tree approximation and with the usual Cayley trees with connectivity $K=2$. 
One can observe that, in both cases, the Cayley-tree approximation does not capture the spatial profile of the superconducting order parameter better than the usual Cayley tree lattice. 
It is also interesting to note that the variation of the order parameter within shells $\sigma$ is comparable to the averaged value of the order parameter, indicating another reason why the Cayley-tree approximation fails in these cases.

\subsection{Numerical solution of the BdG equations}
\label{sec:BdG}

\begin{figure*}[htbp]
  \captionsetup[subfigure]{labelformat=empty}
  \centering
  \begin{tabular}{cc}
    \subfloat[a) $\{6,\,3\}$ lattice]{
      \includegraphics[width=0.45\textwidth]{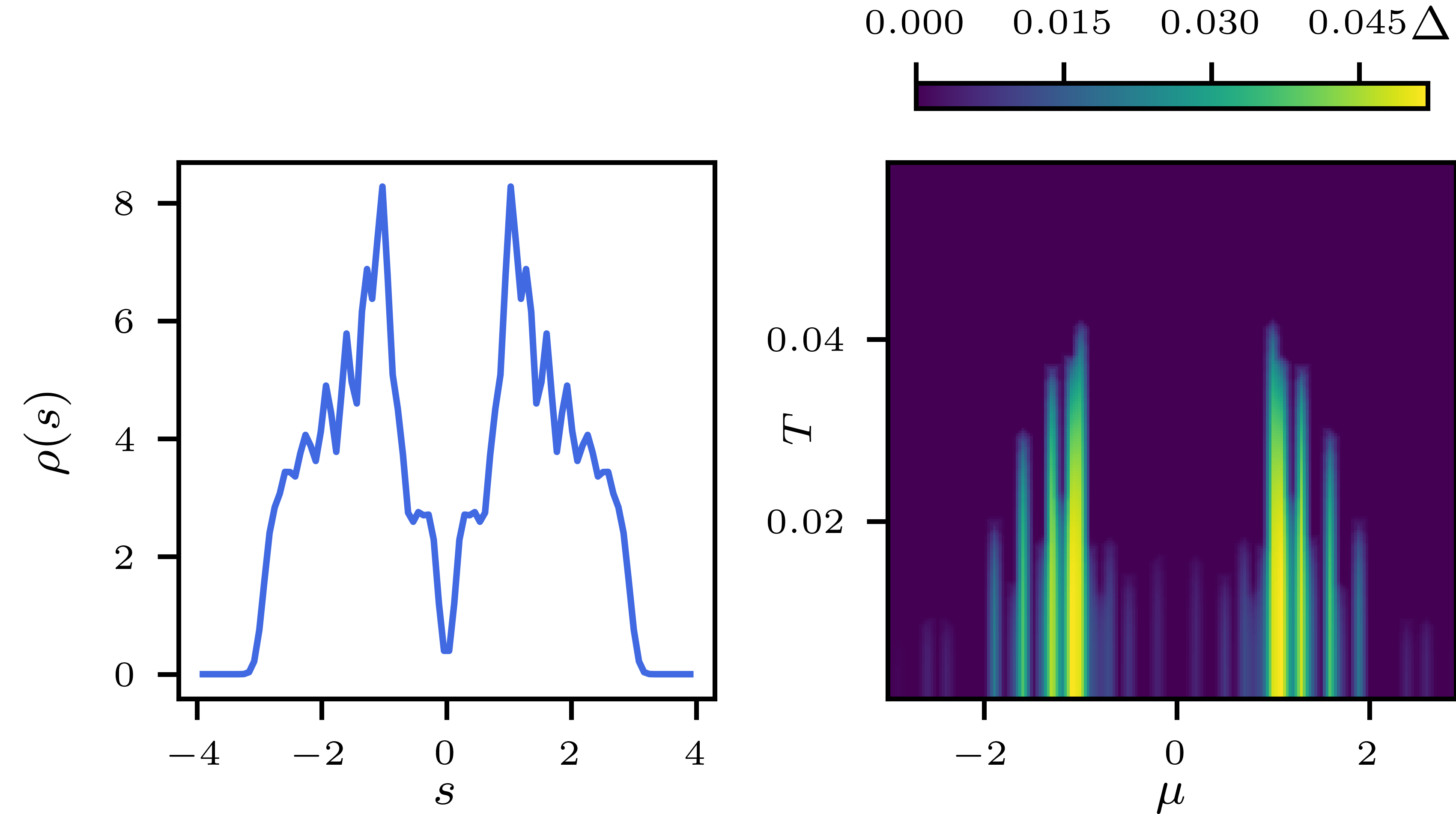}
      \label{fig:p=3_fig1}
    } &
    \subfloat[b) $\{7,\,3\}$ lattice]{
      \includegraphics[width=0.45\textwidth]{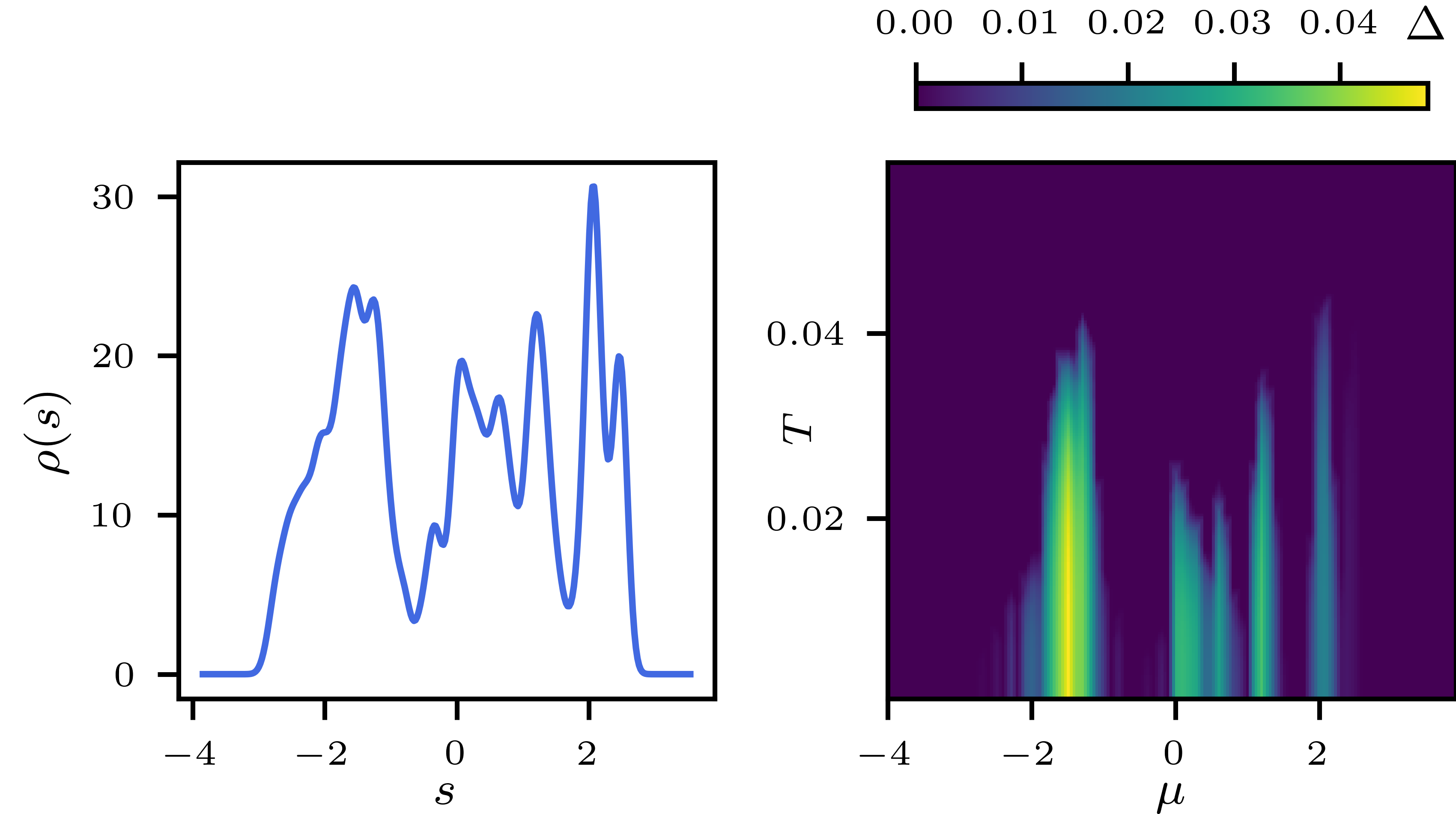}
      \label{fig:p=3_fig2}
    } \\
    \subfloat[c) $\{8,\,3\}$ lattice]{
      \includegraphics[width=0.45\textwidth]{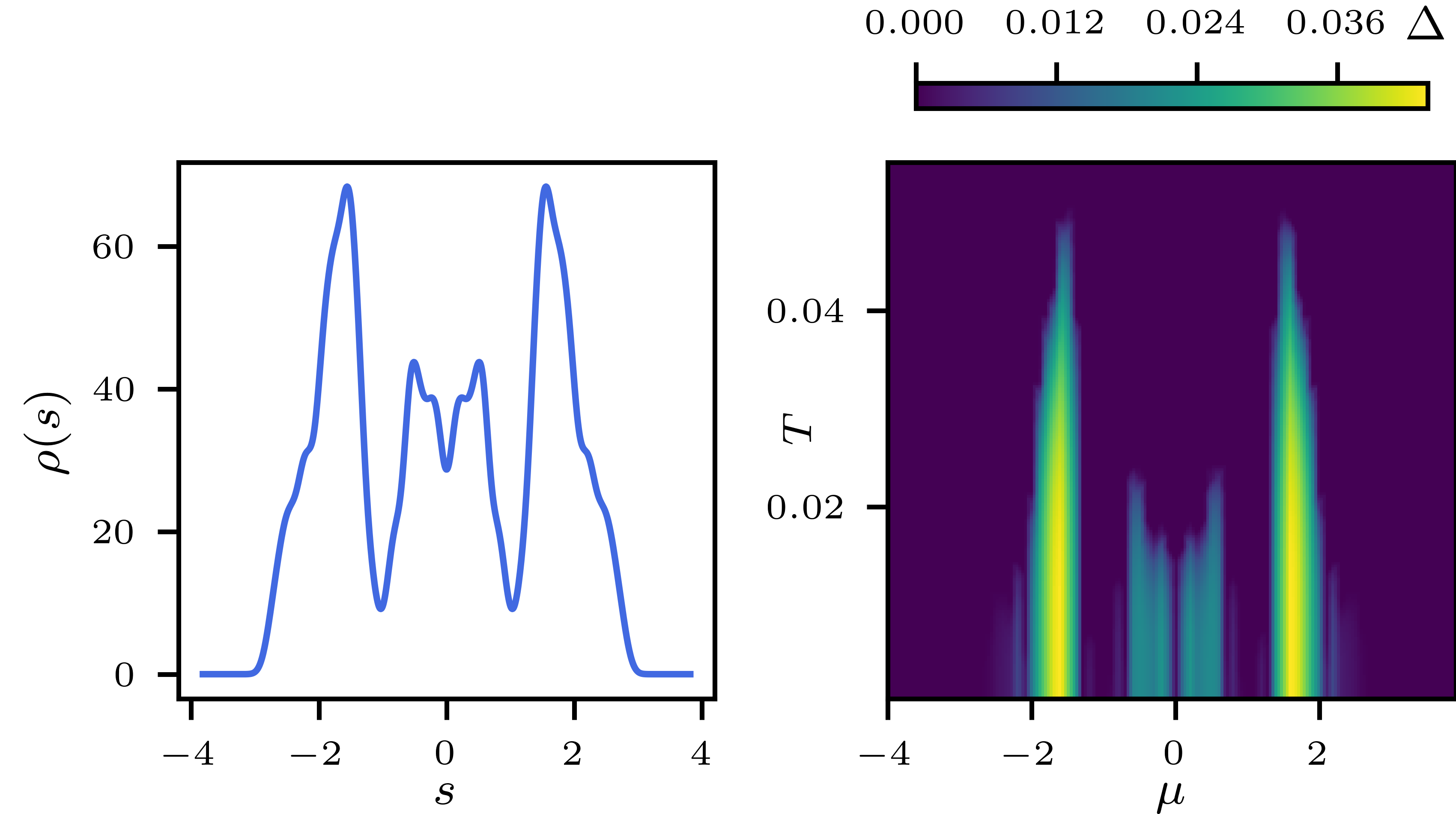}
      \label{fig:p=3_fig3}
    } &
    \subfloat[d) $\{9,\,3\}$ lattice]{
      \includegraphics[width=0.45\textwidth]{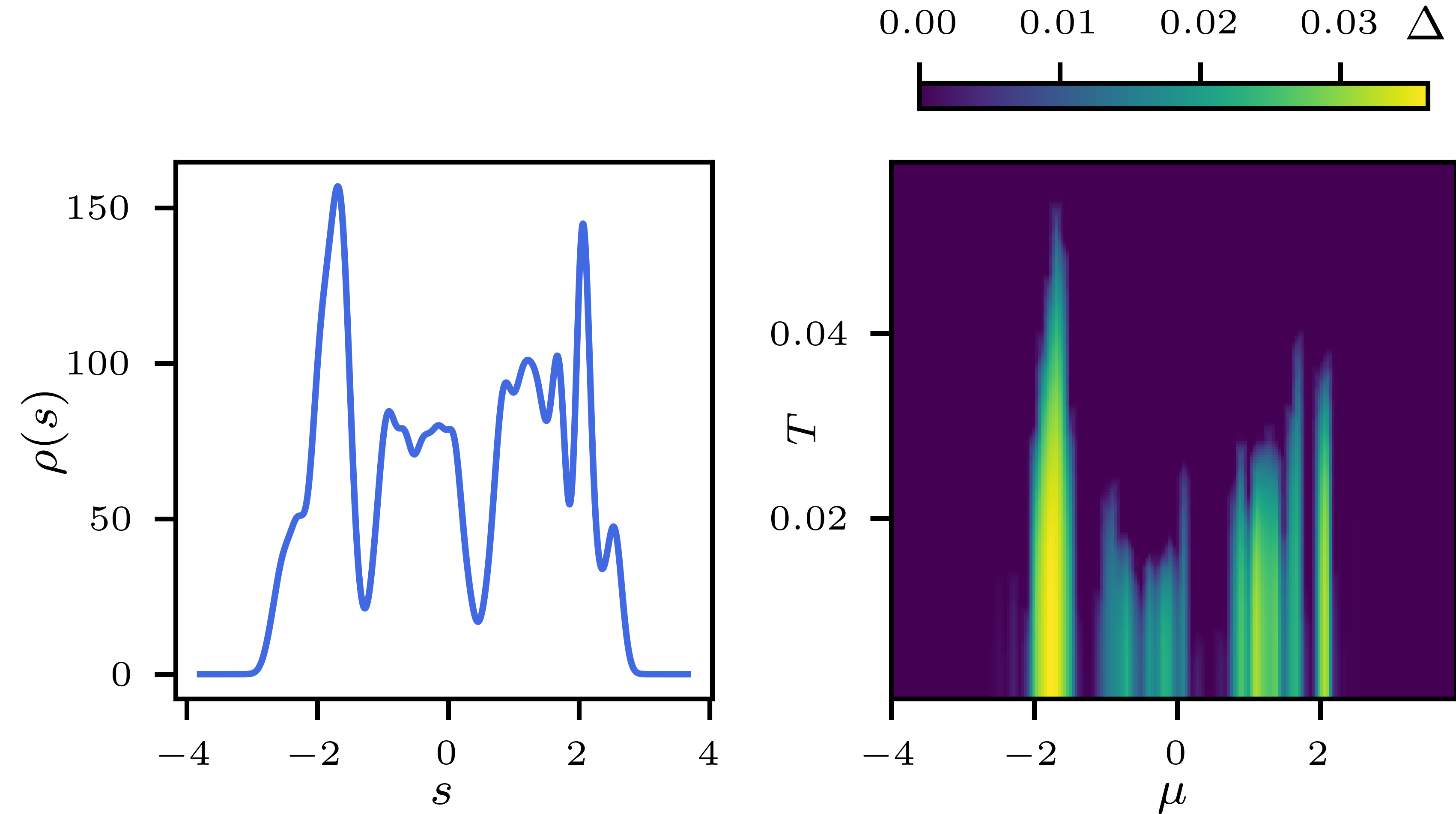}
      \label{fig:p=3_fig4}
    }
  \end{tabular}
  \caption{Densities of states (DoS) (left panels) and $(\mu,\,T)$ phase diagrams (right panels) for the $\{6,3\}$, $\{7,3\}$, $\{8,3\}$, $\{9,3\}$ hyperbolic flakes. The flakes are chosen to be polygon-centered, and contain $l=4$ layers (see the main text for the explanation). The Hubbard coupling constant is set to $U=1$.}
  \label{fig:PD+DoS_q=3}
\end{figure*}

\begin{figure*}[htbp] 

    \captionsetup[subfigure]{labelformat=empty}
  \centering
  
  \subfloat[a) $\{4,\,4\}$ lattice]{
    \includegraphics[width=0.48\textwidth]{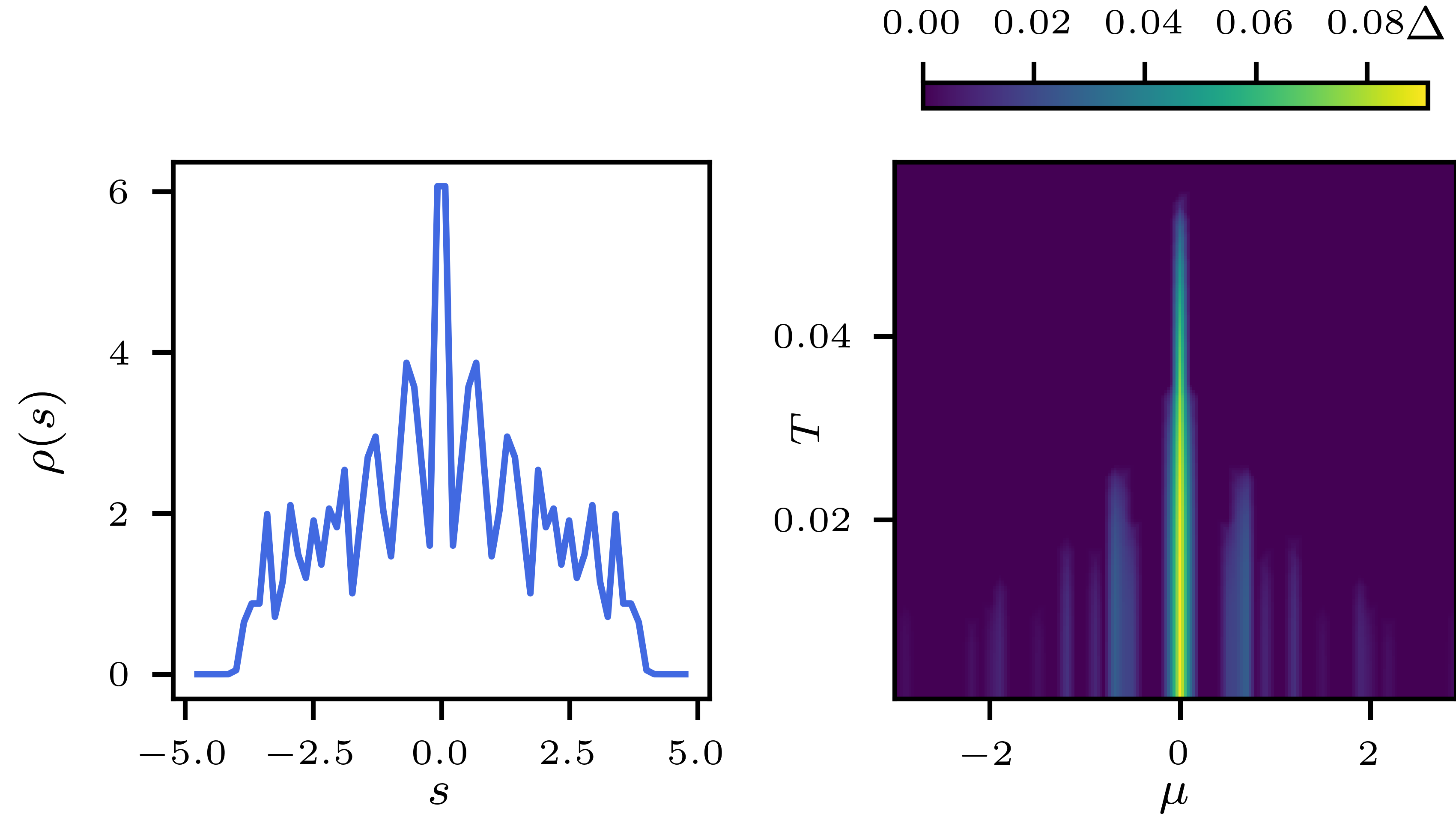}
    \label{fig:fig2}
    }
  \hfill
  \subfloat[b) $\{4,\,5\}$ lattice]{
    \includegraphics[width=0.48\textwidth]{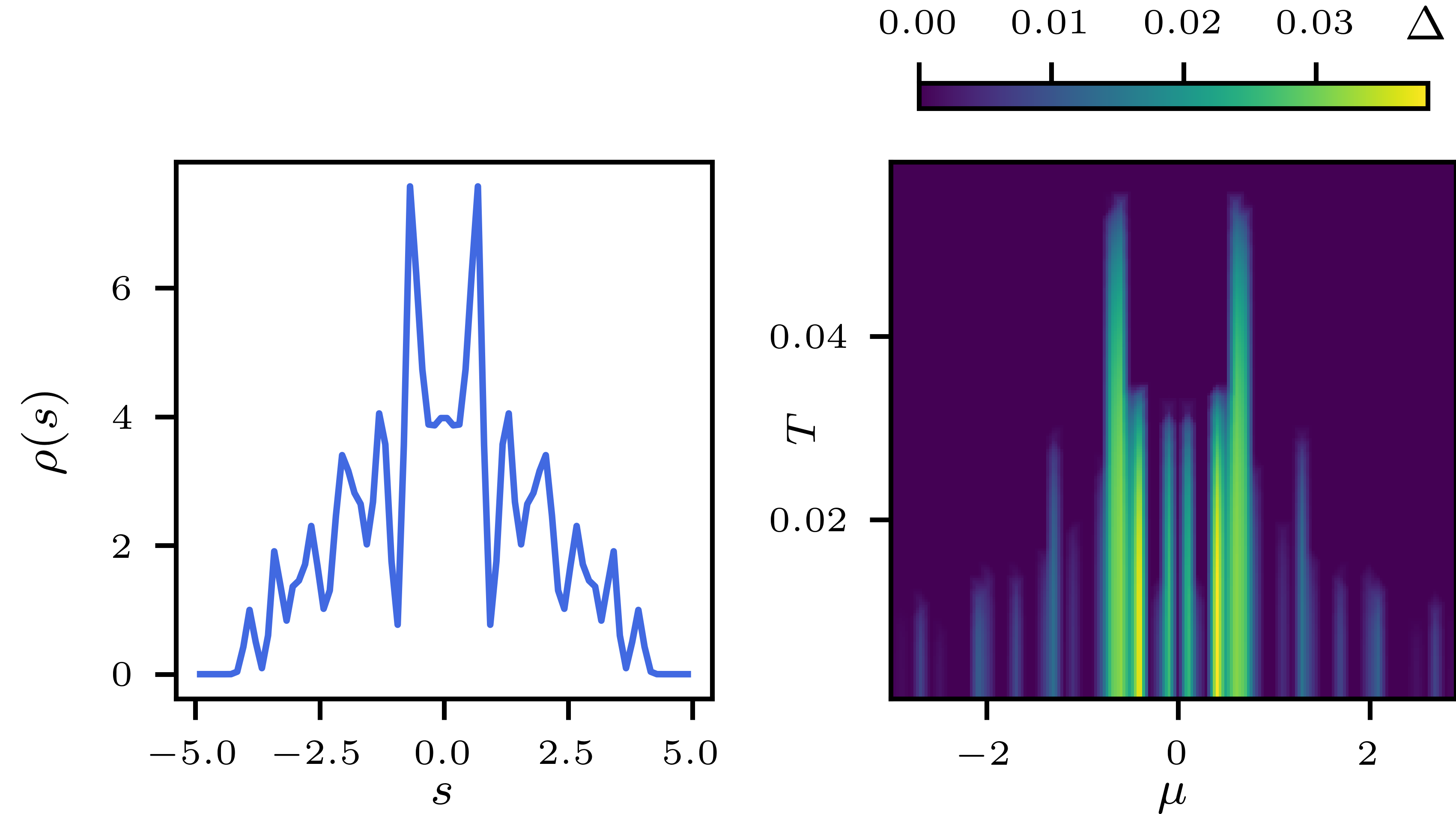}
    \label{fig:fig3}
  }

  \vspace{0cm} 
  \subfloat[c) $\{4,\,6\}$ lattice]{
    \includegraphics[width=0.48\textwidth]{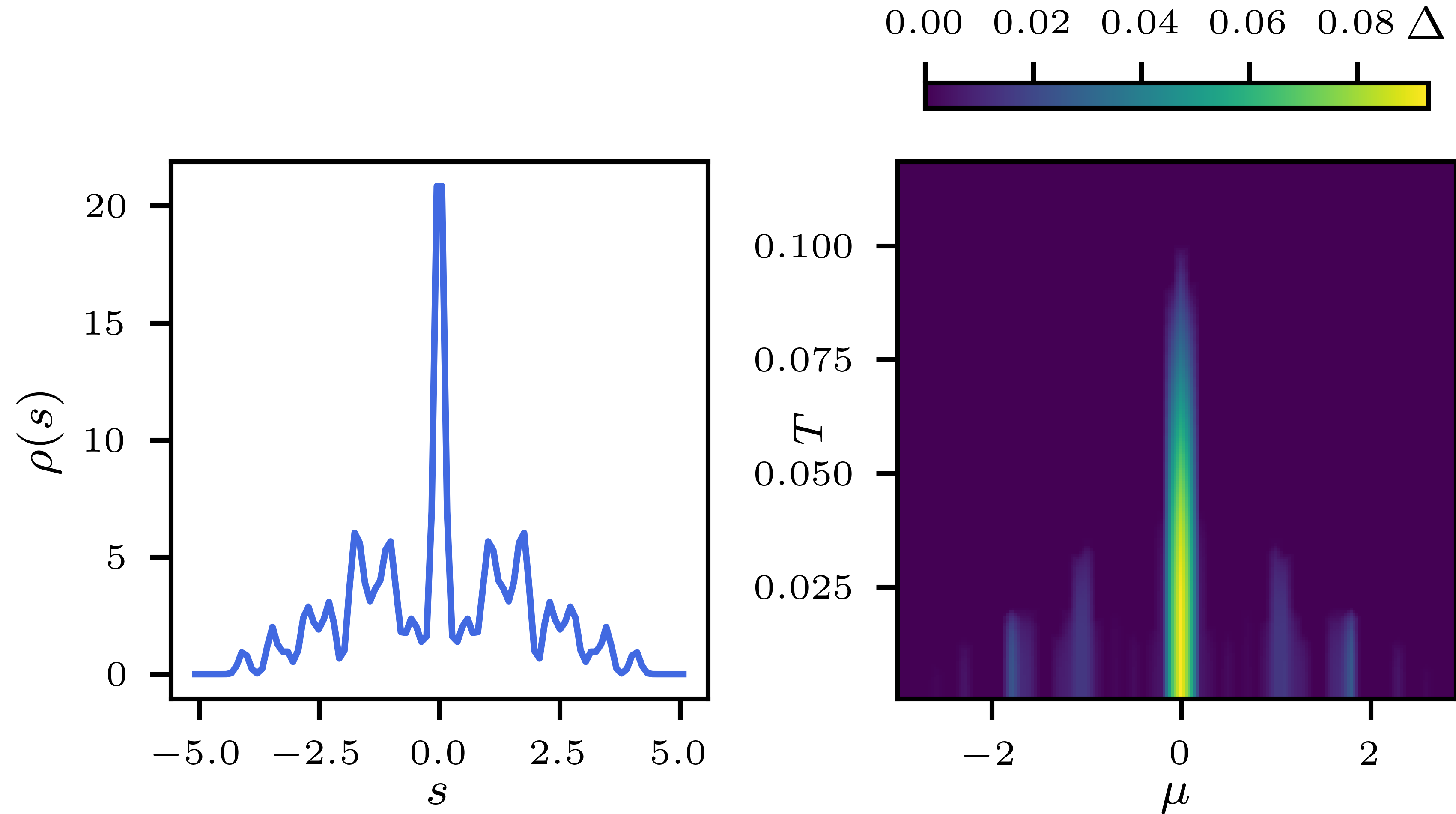}
    \label{fig:fig4}
  }
  \hfill
  \subfloat[d) $\{4,\,7\}$ lattice]{
    \includegraphics[width=0.48\textwidth]{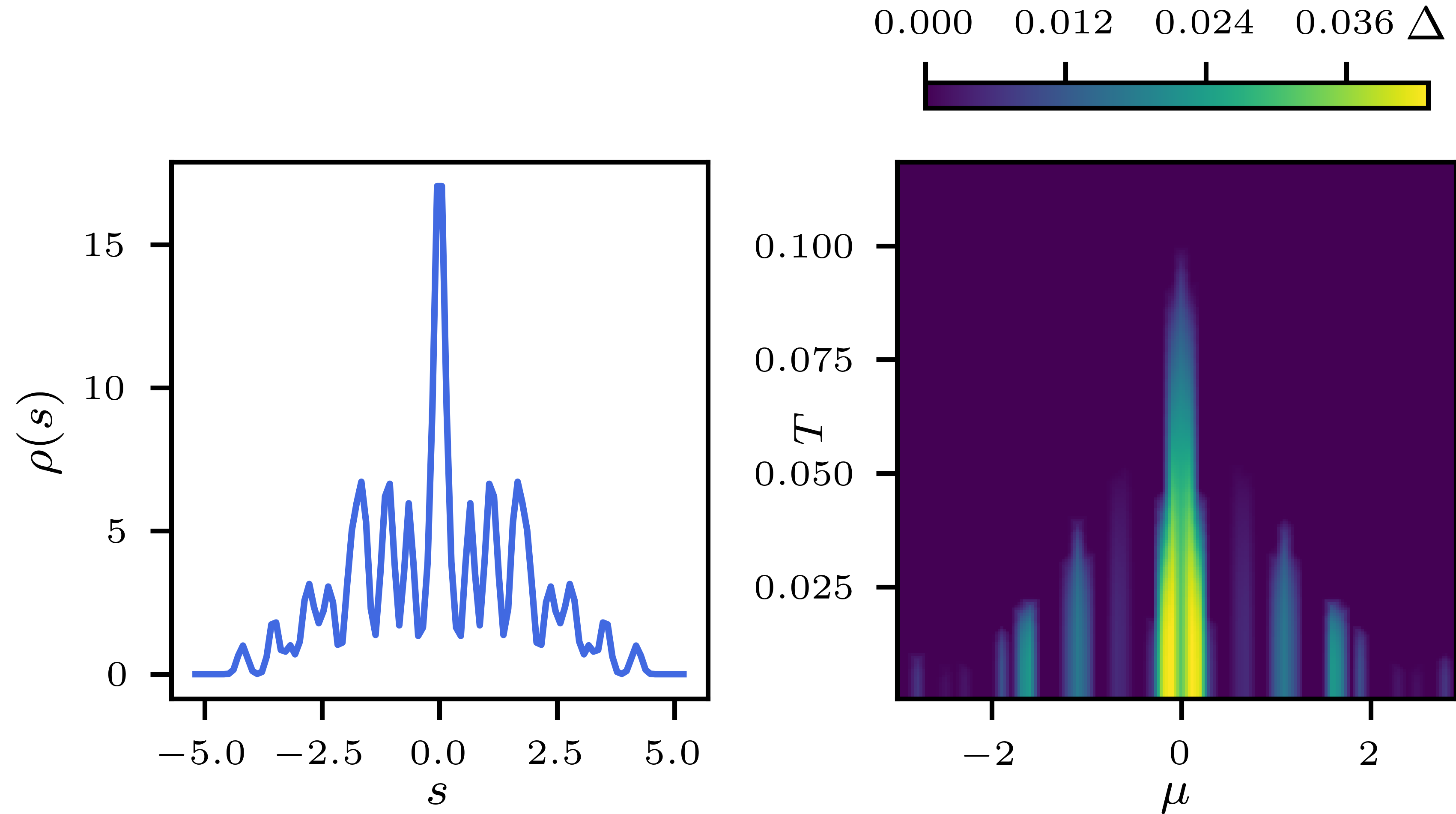}
    \label{fig:fig5}
  }

  \caption{Density of states (DoS) (left panels) and $(\mu,\,T)$ phase diagram (right panels) plotted for the $\{4,\,4\}$, $\{4,\,5\}$, $\{4,\,6\}$, and $\{4,\,7\}$ hyperbolic flakes. The flakes are chosen to be polygon-centered, and contain $l=4$ layers (see the main text for the explanation). The Hubbard coupling constant is $U=1$.}
  \label{fig:PD+DoS_p=4}
\end{figure*}

\begin{figure}[h]
    \centering
    \subfloat[]{%
        \includegraphics[width=0.8\columnwidth]{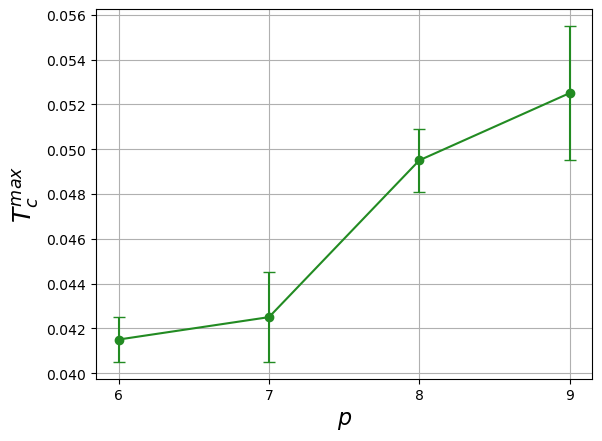}
        \label{fig:T(p)}
    }
    
    \vspace{1em} 
    
    \subfloat[]{%
        \includegraphics[width=0.8\columnwidth]{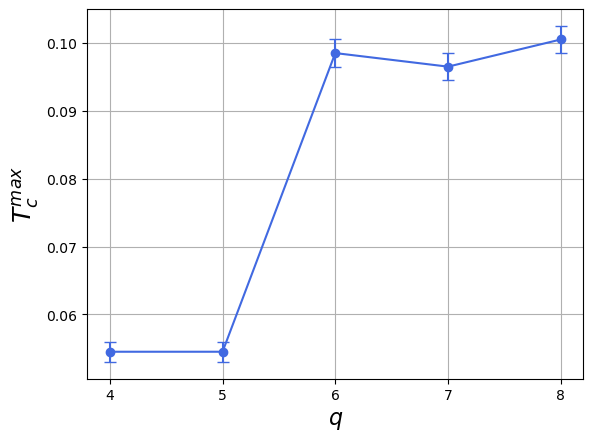}
        \label{fig:T(q)}
    }
    
    \caption{The critical temperature $T_{c}^{max}$, maximized with respect to the chemical potential $\mu$, as a function of (a) $p$ (with $q=3$) and (b) $q$ (with $p=4$).}
    \label{fig:main_figure}
\end{figure}
\begin{figure*}[ht]
    \centering
    \includegraphics[width=0.8\textwidth]{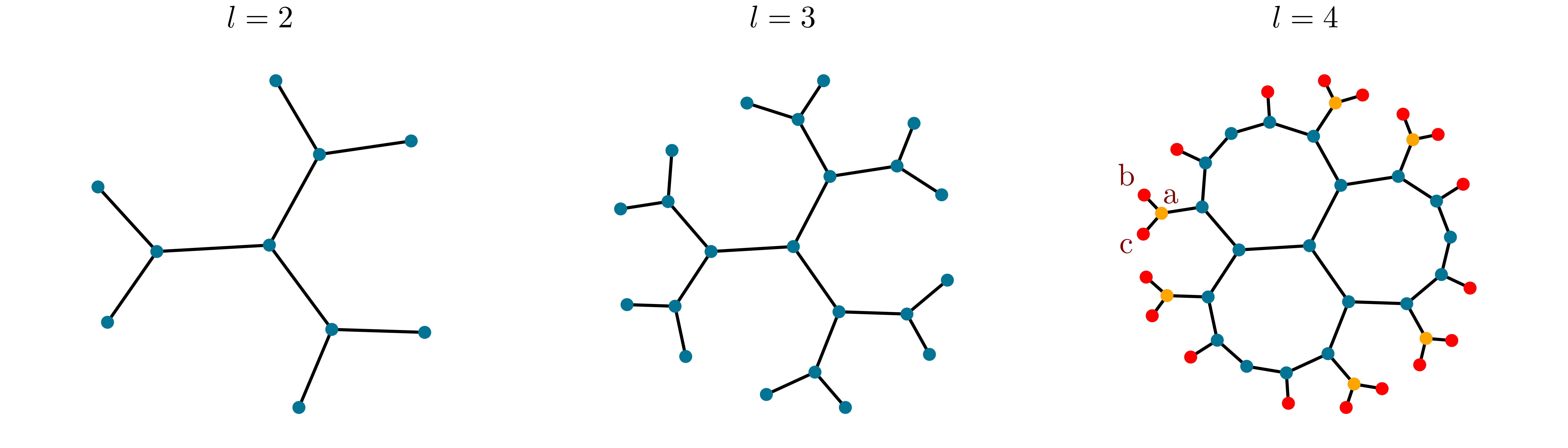}
    \caption{Construction of site-centered $\{8,3\}$ hyperbolic flakes. The starting point ($l=0$) is a single site. 
    At each following step new bonds are attached to each boundary site, and new sites are generated. Starting from the step $l=4$, closed cycles are formed. The last panel also illustrates the dangling sites (red and orange): they may be removed at the final stage of the construction if one aims to obtain    a smooth boundary.}
    \label{fig:site-centered_flake}
\end{figure*}

\begin{figure*}[ht]
    \centering
    \includegraphics[width=0.8\textwidth]{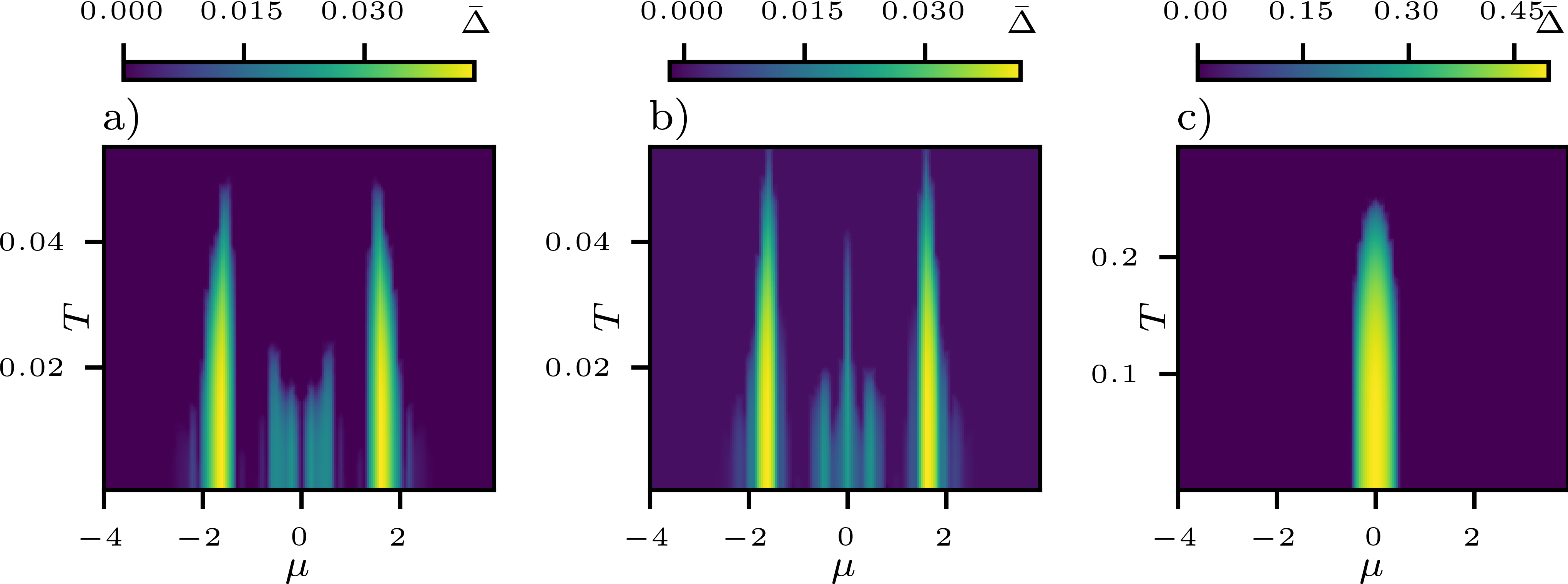}
    \caption{$(\mu,T)$ phase diagrams for finite patches of the $\{8,3\}$ lattice. a) polygon-centered flake; b) site-centered flake with smooth boundary conditions; c) site-centered flake with dangling sites at the boundary.}
    \label{fig:three_PDs}
\end{figure*}

The semi-analytical approaches explored in the previous section are limited to specific cases of hyperbolic lattice geometries (even $p$) and to a particular value of the chemical potential ($\mu = 0$). 
While unveiling boundary superconductivity in hyperbolic lattices, they do not give access to the full parametric space of the model. 
To overcome these limitations and handle arbitrary chemical potentials and general $\{p,q\}$ hyperbolic lattices without relying on the approximate radial symmetry, we perform direct numerical solutions of the BdG equations on finite hyperbolic flakes. 
Although this approach does not yield closed-form solutions, it allows us to identify phenomena missed by the tree approximation and obtain a complete phase diagram of the attractive Hubbard model in hyperbolic spaces by systematically varying the geometric parameters $p$ and $q$ as well as the chemical potential~$\mu$. Here, we focus less on the distinction between the bulk vs.~boundary superconductivity, but  pay more attention to the role of parameters and boundary conditions in shaping the phase diagram of the model and defining its critical temperature.

As in the previous section, we solve numerically the self-consistent system of Eqs.~\eqref{eq:BdG_Ham} and~\eqref{eq:selfcons_delta}, with the coupling parameter set to $U=1$.
Unlike the previous section, we will here construct hyperbolic flakes centered on an elementary cell (a $p$-sided polygon)
and generate further layers by applying the hyperbolic translation between adjacent polygons. 
More precisely, we start from a single seed polygon and translate it across its edges to the adjacent polygons. 
In the next step, we apply the same procedure to each polygon found in the system, and so on.\footnote{We note that this construction is slightly different from the site-inflation, discussed
in Ref.~\cite{Jahn:2020}, where at each step one includes all the polygons adjacent to the existing ones either across an edge \emph{or a site}.} 
Such construction results in a smooth boundary condition without dangling bonds, and we will be considering systems obtained through four steps of such an iterative procedure. To investigate the role of the boundary termination, for the $\{8,3\}$ lattice we further explicitly compare results for the flake described above to a flake constructed through accretion of sites with growing graph distance from a seed site in the spirit of the shells $S_l$ from Sec.~\ref{sec:Cayley_tree_apx}. 
This alternative construction results in a ragged boundary termination with numerous dangling bonds, which we observe to have drastic implications for the obtained phase diagram.

An obvious advantage of dealing with hyperbolic lattices is that one can vary independently the number of polygon edges 
$p$ and the site degree $q$. Our first set of numerical experiments concerns the behaviour with respect to the variation of $p$ while keeping $q$ fixed (we set it to $q=3$). 
This family starts from the (flat-space) hexagonal lattice $\{6,3\}$, and goes all the way to the Cayley tree case $\{\infty, 3\}$.

The phase diagrams corresponding to the first four instances ($p=6,\ldots,9$) are shown in the right panels of Fig.~\ref{fig:PD+DoS_q=3} (a--d): 
the values of the gap\footnote{Here we choose to plot $\text{min}\,\Delta$, where the minimization is taken over the lattice sites.} $\Delta$ are
plotted as a function of temperature $T$ and chemical potential $\mu$. 
As the first observation, we note that the form of the superconducting region is strongly correlated with the density of states for the corresponding tight-binding models, shown in the left panels of Fig.~\ref{fig:PD+DoS_q=3}(a--d). 
This is what one expects to find within the BdG framework in a homogeneous system \cite{BdG_book}; however, it does not follow automatically for non-homogeneous systems such as finite hyperbolic flakes, where the boundary is expected to give important (if not dominant) contribution to the density of states. 
One can also notice that for $p$ even, both the phase diagrams and the DoS plots are symmetric with respect to $\mu = 0$, which is a consequence of the chiral symmetry of the Hubbard model on a bipartite lattice.

In the second series of numerical experiments we consider the variation of the coordination number $q$, while fixing the polygons to be squares ($p=4$).
The corresponding $(T,\mu)$ phase diagrams as well as the relevant DoS are shown on Fig.~\ref{fig:PD+DoS_p=4}. 
The plots are symmetric, since $p=4$ corresponds to bipartite lattices, and we observe again strong correlation between the DoS and the phase diagrams. 
Another interesting feature that can be noticed in Fig.~\ref{fig:PD+DoS_p=4} is the subtle dependence on the parity of $q$: while for even $q$ the maximal value of $T_c$ is reached for $\mu=0$, for odd $q$ the point $\mu=0$ corresponds to a local minimum of $T_c$.\footnote{Sometimes, like in the case of $\{4,7\}$, artifacts of the smoothening procedure lead to a mild misalignment of the plot, visually shifting the local minimum slightly away from $\mu=0$. 
Relatedly, the apparent maximum of DoS at $\mu=0$ is also caused by the smoothening procedure. Exact diagonalization shows complete absence of zero modes.}

A quantity of special interest is the superconducting temperature $T_c$: we may assume tuning the chemical potential such that $T_c$ is maximized, and call the result $T_c^{max}$. Although Fig.~\ref{fig:T(p)} demonstrates only a mild variation of $T_c^{max}$ with respect to the variations of $p$, Fig.~\ref{fig:T(q)} reveals a more pronounced behavior when $q$ is modified: we observe a sudden jump while passing from $q=5$ to $q=6$. 
This can be explained by observing that 
a collection of energy levels 
near zero energy, seen in the DoS (or phase diagram) for the $\{4,5\}$ lattice [see Fig.~\ref{fig:PD+DoS_p=4}(b)], accumulates even closer near $\mu=0$ 
for higher values of $q$. This is in contrast with what happens in Fig.~\ref{fig:PD+DoS_q=3}: the brightest bands tend to repel from the origin when we increase $p$, and so do not enhance each other. 

Finally, we would like to touch upon the question of how different the two constructions of finite hyperbolic flakes discussed in this paper are. Recall that one may consider building a finite hyperbolic flake: starting from either a central site while preserving $\mathbb{Z}_q$ rotation symmetry, or from a central polygon while preserving $\mathbb{Z}_p$ rotation symmetry.\footnote{Taking also reflections into account, the symmetry groups are found to be the dihedral groups $\mathbb{D}_q$ and $\mathbb{D}_p$, respectively.} As it was already pointed out, iterative construction of a site-centered flake (see Fig.~\ref{fig:site-centered_flake}) follows closely the Cayley tree construction, with the difference that sometimes the branches connect, forming cycles. In this construction, one naturally gets a lot of dangling sites at the boundary which are connected to just a single neighboring site. In the following discussion of site-centered hyperbolic flakes, we will explicitly consider two distinct
possibilities: the first is to keep the dangling sites as part of the flake while keeping the boundary ragged, and the second option is to remove the dangling sites while leaving a smoother boundary.

For the sake of a simple comparison, we will consider finite patches of the $\{8,3\}$ lattice. 
The polygon-centered flake is obtained by accreting three more layers of polygons around the central one: this gives in total 768 sites, and the most remote sites are located at the radius $r_{max}\simeq 5.42R$ from the center. 
By virtue of our construction, the polygon-centered flake has no dangling sites, i.e., it has a smooth boundary.
To construct the site-centered flake, we apply 10 iterations, which gives 757 sites, with the most remote ones located at the distance $r_{max}\simeq 5.72R$. 
Finally, for the site-centered flake with removed dangling sites we apply 11 iterations 
followed by the removal of the one-neighbor sites: this gives 640 sites, and the maximal distance from the center is $r_{max}\simeq 5.52R$. 
Therefore, we consider three flakes with comparable numbers of sites and a similar geometric extent.
The three choices of flakes allow us to investigate the role of centering (polygon vs.~site) and of the boundary termination (smooth vs.~with dangling sites).

The phase diagrams for these three cases are shown in Fig.~\ref{fig:three_PDs}. We observe that the polygon-centered flake and the site-centered flake with smooth boundary exhibit similar structures of the phase diagram as well as comparable maximal critical temperatures. 
A noticeable difference can be found near $\mu=0$, where in the first case superconductivity is absent, while in the second case we find a peak of $T_c$. 
In contrast, the case of the site-centered flake with dangling sites at the boundary demonstrates a sharp difference with respect to the first two cases: we see a well-pronounced band near $\mu=0$, showing critical temperature about five times larger than before. 
This enhancement of $T_c$ can be traced to the altered boundary termination. 
Namely, one can find sites close to the boundary [orange in Fig.~\ref{fig:site-centered_flake}(c) 
that are neighboring two dangling sites ([red in Fig.~\ref{fig:site-centered_flake}(c)]); an example of such a triple in Fig. ~\ref{fig:site-centered_flake}(c) is labeled by  $a$, $b$, and $c$. Then one can associate a state
\begin{equation}
\label{eqn:dangling-triplet-states}
    (c^{\dagger}_{b,\,\sigma}-c^{\dagger}_{c,\,\sigma})|0\rangle.
\end{equation}
with every such triplet,\footnote{The symmetry of a site-centered flake with dangling sites is enhanced to $(\mathbb{Z}_2)^k\rtimes \mathbb{D}_3$, where $k$ is the number of triplets, and the $\mathbb{Z}_2$ factors swap the $a$ and $b$ sites of a chosen triplet, while leaving all the other sites intact. The states defined by Eq.~(\ref{eqn:dangling-triplet-states}) are odd under one of $k$ $\mathbb{Z}_2$ transformations.} 
and every such state can be shown to be an exact zero-energy eigenstate
of the hopping part of the Hamiltonian. It is the proliferation of these zero modes that is responsible for the dramatic enhancement of superconductivity near $\mu=0$.

\section{Ginzburg-Landau description}
\label{sec:ginzburg-landau}
In the previous chapter, we analyzed superconductivity on hyperbolic lattices using both analytical approximations and direct numerical solutions of the BdG equations. This lattice-based perspective is well suited to capturing discrete structures and boundary effects, and to gain a broader understanding, it is natural to complement the lattice analysis with a continuum description. 

In the following, we construct a Ginzburg–Landau (GL) theory on the Lobachevsky plane, i.e., a constant-time slice of the AdS space. This approach allows us to focus on generic features of hyperbolic geometry, without being tied to lattice-specific details such as the density of states of a concrete tiling. 
In particular, we will test whether key phenomena observed in the lattice setting, such as the enhancement or suppression of superconductivity near boundaries, also emerge in a continuous formulation. 
By studying both the infinite-space limit and finite clusters with boundaries, the continuum framework provides a complementary viewpoint that clarifies which effects are intrinsic to hyperbolic geometry itself and which are specific to discrete structures.

We therefore consider the GL free energy functional
\begin{equation}
    \mathcal{F}=\int d^2z\,\sqrt{g}\,\left(\partial_{a}\bar{\phi}\partial^a\phi\,+\,V(\phi)\right),
\end{equation}
where $\phi$ is the order parameter, approximately proportional to the $s$-wave gap $\Delta$ for temperatures close enough to $T_c$ \cite{Gor'kov:59}, and $g_{ab}$ is the metric on the hyperbolic plane. 
Our starting point is the Poincaré disk representation with coordinates
$(r,\,\varphi)$ where $\varphi\in[0, 2\pi)$ and $r\in[0,\,\infty)$. 
The limit $r\rightarrow\infty$ then corresponds to the conformal boundary of the hyperbolic plane, while $r\rightarrow0$ is the deep interior. 
The metric is 
\begin{equation}\label{Disc_Metric}
    g_{ab}=\left( 
    \begin{matrix}
        1 && 0\\
        0 && R^2\sinh^2\tfrac{r}{R}
    \end{matrix}
    \right),
\end{equation}
where $R$ is the radius of curvature of the hyperbolic plane. The potential is chosen to be of the simplest form,
\begin{equation}
\label{eqn:mexican-potential}
    V(\phi)=-\tau |\phi|^2 + \frac{1}{4}|\phi|^4,
\end{equation}
and $\tau\propto T_c-T$. 
Evidently, for $\tau<0$, the theory is in the normal conducting phase, while for $\tau>0$ spontaneous symmetry breaking takes place, and the order parameter develops a vacuum expectation value. Technically, we will consider inhomogeneous solutions of the equation of motion 
\begin{equation}\label{AdS_Scalar_EOM}
    \frac{1}{\sqrt{g}}\partial_a\left(\sqrt{g}g^{ab}\partial_b \right)\phi-\frac{dV(\phi)}{d\phi}=0,
\end{equation}
augmented with certain boundary conditions.
Generally, we are interested in three cases: the entire Poincar\'e disc ($r=0\dots\infty$), a disc with finite boundary cutoff ($r=0\dots r_1$), and an annulus such that the boundaries of the annulus are concentric circles (i.e., $r=r_0\dots r_1$). Eq.\eqref{AdS_Scalar_EOM} in the Poincar\'e disc representation is too complicated to allow for analytical insights. Hence, we adopt the following strategy. In Sec.~\ref{sec:GL-semi-infinite}, we analyze GL theory on the hyperbolic half-plane, where the equations of motion are amenable to semi-analytical approach based on phase portrait analysis, and obtain condensate profiles corresponding to different $\tau$ and $R$. The geometries we are dealing then do not directly correspond to the aforementioned finite-radius disc and annulus, but at the end of the section we will establish connections between the considered cases.

\subsection{Ginzburg-Landau theory on the hyperbolic half-plane}
\label{sec:GL-semi-infinite}

In this discussion, we adopt the Poincaré half-plane model of the hyperbolic plane, which we parameterize with the coordinates $-\infty<\rho<+\infty$, $-\infty<x<+\infty$. Here, $x$ is the coordinate along the conformal boundary and $\rho$ is the interior direction such that $\rho \to +\infty$ ($\rho \to -\infty$) corresponds to the conformal boundary (to the deep interior).
We will later narrow our attention to the submanifold corresponding to the finite range $\rho \in (-\infty,\rho_0)$ with finite $\rho_0$\footnote{In the Poincare disc coordinates, this region would correspond to a horodisc -- a disc bounded by a horocycle, and not directly to a disc centered at $r=0$ in the coordinates of \eqref{Disc_Metric}.}
In these coordinates,
\footnote{Recall that one of the standard coordinates in the Poincaré half-plane model are $x \in (-\infty,+\infty)$ along the holographic boundary and $y\in(0,+\infty)$ perpendicular to the holographic boudnary, in which case $ds^2 = (dx^2 + dy^2)R^2/y^2$. Our adopted choice of coordinates is obtained by keeping $x$ while substituting $y = 2e^{-\rho/R}$. Note that increasing $\rho$ corresponds to \emph{decreasing} $y$, i.e., the holographic boundary at $y\to 0$ translates to $\rho\to +\infty$. Horodisks correspond to regions with $y>y_0$, which in our coordinates translates to $\rho < \rho_0$.}
the metric takes the form
\begin{equation}\label{HalfPlaneMetric}
    g_{ab}=\left( 
    \begin{matrix}
        1 && 0\\
        0 && \frac{R^2}{4} e^{2\rho/R}
    \end{matrix}
    \right).
\end{equation}
Looking for profiles depending only on $\rho$, and using the metric \eqref{HalfPlaneMetric}, Eq.~\eqref{AdS_Scalar_EOM} can be written more explicitly as
\begin{equation}
\label{eqn:GL-in-half-plane}
    \phi''+\frac{1}{R}\phi'+\tau\phi-\frac{1}{2}|\phi|^2\phi=0,
\end{equation}
where the prime indicates the derivative with respect to $\rho$. 
This is an autonomous differential equation, and one can readily analyze qualitative behavior of the solutions. We will parameterize the order parameter as $\phi=q\,e^{i\alpha}$ with real parameters\footnote{It is convenient to allow $q$ to take negative values, so that the range of the phase is $\alpha\in[0,\pi)$. In fact, later we will consider only the case $\alpha=0$.} $q$ and $\alpha$, and assume that the phase remains constant. 
Introducing the ``momentum'' $p=q'$, we can rewrite the equation~as
\begin{align}\label{eq:dynamical_system}
\begin{split}
    q'&=p,\\
    p'&=-\frac{1}{R}p-\tau\,q+\frac{1}{2}q^3.
\end{split}
\end{align}
The analysis of possible profiles of the order parameter therefore boils down to the study of the dynamical system~\eqref{eq:dynamical_system} with the coordinate and momentum $(q, p)$, with $\rho$ playing the role of the ``time'' variable.

Before we proceed, let us observe that the term related to curvature plays the role of damping in the dynamical system at hand. However, in the following, we will be specifying boundary conditions at the external boundary (i.e., at $\rho\to+\infty$), and search for the corresponding profile extending in the bulk. This corresponds to moving backward in the ``time'' direction $\rho$; therefore, the damping is effectively negative and leads to destabilization.

We now continue with the analysis, focusing on the case $\tau>0$ corresponding to the presence of finite superconducting order parameter in the deep interior. 
In order to proceed, it is natural to distinguish two different cases.


\subsubsection{Poincaré half-plane for $4R^2\,\tau<1$} 

The system has three fixed points: 
\begin{equation}
(q_0,\,p_0)=(0,\,0)\quad\textrm{and}\quad (q_0,\,p_0)=(\pm\sqrt{2\tau},\,0),
\end{equation}
corresponding to the normal conductor phase and to the superconductor phase, respectively.
The exponents at the first fixed point are found to be
\begin{equation}
\lambda_{1,2}=\frac{1}{2R}\left(-1\pm\sqrt{1-4R^2\tau}\right),
\end{equation}
which are both real and negative; therefore, this is a stable fixed point (a stable node). 
At the other two fixed points, the exponents are given by 
\begin{equation}
\lambda_{1,2}=\frac{1}{R}\left(-1\pm\sqrt{1+8R^2\tau}\right),
\end{equation}
which are real numbers with opposite sign; therefore, these fixed points are saddles.

\begin{figure}[t]
    \includegraphics[width=0.42\textwidth]{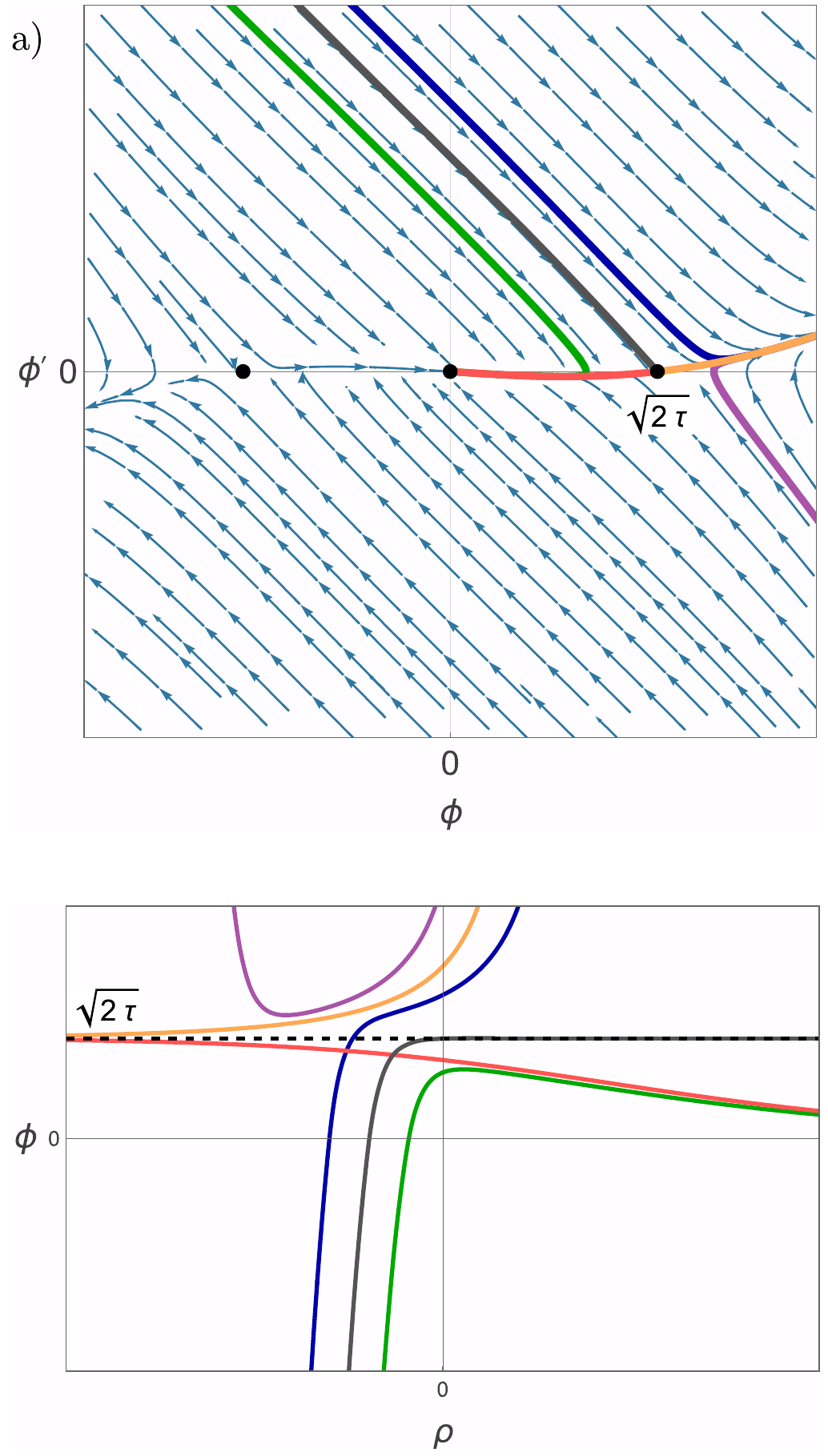}
    \caption{An illustration of the (a) phase trajectories and (b) spatial
    profiles of the order parameter in the Poincar\'e half-plane for $4R^2\tau<1$ (here we set $R=1$ and $\tau = 1/16$). See the text for more comments on the trajectories.}
    \label{fig:GLPhase1}
\end{figure}

  
Let us now discuss some types of trajectories, starting from the case of the entire infinite half-plane. 
Besides the uniform solution $\phi = \sqrt{2\tau}$, there is a second trajectory such that the order parameter is finite along it, and it corresponds to a profile interpolating between the unbroken vacuum $\phi_u=0$ at the conformal boundary $\rho\rightarrow+\infty$ and a symmetry breaking vacuum (which we fix to be $\phi_b=\sqrt{2\tau}$) in the interior\footnote{This is in direct analogy with domain wall solutions, which in the holographic setting are dual to renormalization group flows on the boundary.} $\rho\rightarrow-\infty$: it is depicted in Fig.~\ref{fig:GLPhase1} by the red line. The order parameter changes monotonically along this trajectory.

We may also choose to impose boundary conditions not at the conformal boundary but at some finite distance $\rho_0$. The requirement of regularity in the interior is still rather restrictive, but now we have three basic choices. 
Two solutions follow automatically from the solutions in the infinite bulk. 
Namely, we can either have a segment of the constant solution [black dashed line in Fig.~\ref{fig:GLPhase1}(b)] if we impose $\phi(\rho_0) = \phi_b$, or a segment of the previously discussed interpolating trajectory (red in Fig.~\ref{fig:GLPhase1}) if $\phi(\rho_0)<\phi_b$. 
The latter case can be interpreted as boundary-suppressed superconductivity. 
In addition, the presence of the boundary enables a third solution, which corresponds to landing on the other saddle's repulsive direction (orange in Fig.~\ref{fig:GLPhase1}). 
In this additional case, $\phi(\rho_0) > \phi_b$, corresponding to boundary-enhanced superconductivity. 
The situation when the order parameter $\phi$ increases toward the boundary 
is in agreement with the results in Sec.~\ref{sec:uni_lat}, where the same type of behavior has been found.

\subsubsection{Poincaré half-plane for $4R^2\,\tau>1$} 

Next, we lower the temperature even further.\footnote{At this point it is natural to question the validity of Ginzburg-Landau description, which is supposed to be faithful only for temperatures sufficiently close to the phase transition point. We will not address this important question and
assume that this approach is still capable of capturing the relevant physics.} Note that from the point of view of physics in the hyperbolic plane, the condition $4R^2\tau=1$ corresponds to the Breitenlohner-Freedman bound~\cite{Breitenlohner:1982jf}.\footnote{See also Ref.~\citenum{Basteiro:2022pyp} for a recent discussion of the Breitenlohner-Freedman bound in the context of hyperbolic tilings.} 
The phenomena beyond the bound are less discussed since the hyperbolic space is often considered in the context of the anti de Sitter/Conformal field theory correspondence, and crossing the bound implies instability in the bulk and violation of unitarity/reflection positivity on the CFT side. However, these issues are not relevant in the present study, and the regime is physically valid.

\begin{figure}[t]   
    \includegraphics[width=0.42\textwidth]{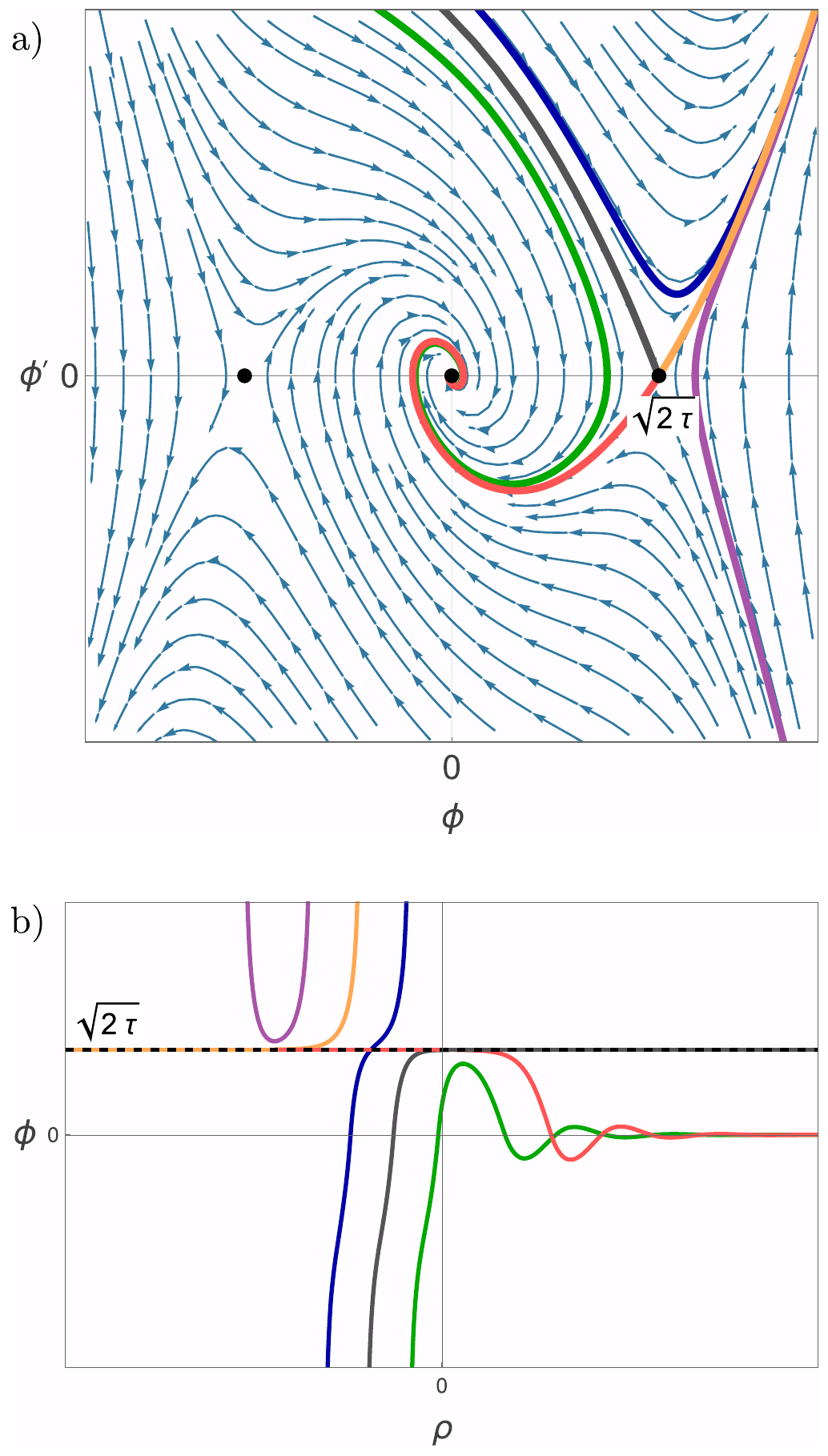}
    \caption{An illustration of the (a) phase trajectories and (b) spatial profiles of the order parameter in the Poincar\'e half-plane for $4R^2\tau>1$ (here we set $R=1$ and $\tau = 2$). See the text for more comments on the trajectories.}
    \label{fig:GLPhase2}
\end{figure}

Performing the same steps as before, we observe that the main difference compared to the discussion above is that the fixed point $(q_0,\,p_0)=(0,\,0)$ changes its nature: the corresponding exponents $\lambda_{1,2}=\frac{1}{2R}\left(-1\pm i\sqrt{4R^2\tau-1}\right)$ are now complex numbers, and so this is a stable focus. 
Starting again from the global case, we find that there are still two solutions. While the constant solution $\phi(\rho) = \phi_b$ is qualitatively the same as before, we find that the interpolating solution connecting $\phi=\phi_b$ at $\rho\to-\infty$ to $\phi=0$ at $\rho\to+\infty$ (which resembles superconducting condensate leaking into emergent metallic phase) now exhibits oscillations rather than being monotonic (red curve in Fig.~\ref{fig:GLPhase2}).\footnote{Holographically, this can be considered as an evidence of the breakdown of RG flow monotonicity, coming along with the violation of unitarity.} In some regard, it could be understood as manifestation of two types of superconductivity set by the ratio of the correlation length (encoded in $\tau$) and the curvature radius $R$, with oscillations and nodal lines of the condensate being a conceptual analog of vortices in type-II superconductors even in the absence of external magnetic field. However, the important difference is that these oscillations occur only within the exponentially decaying part of the profile that can be regarded as a metallic phase emerging upon approaching the holographic boundary (spatial infinity). 

\begin{figure}[t]
    \includegraphics[width=0.42\textwidth]{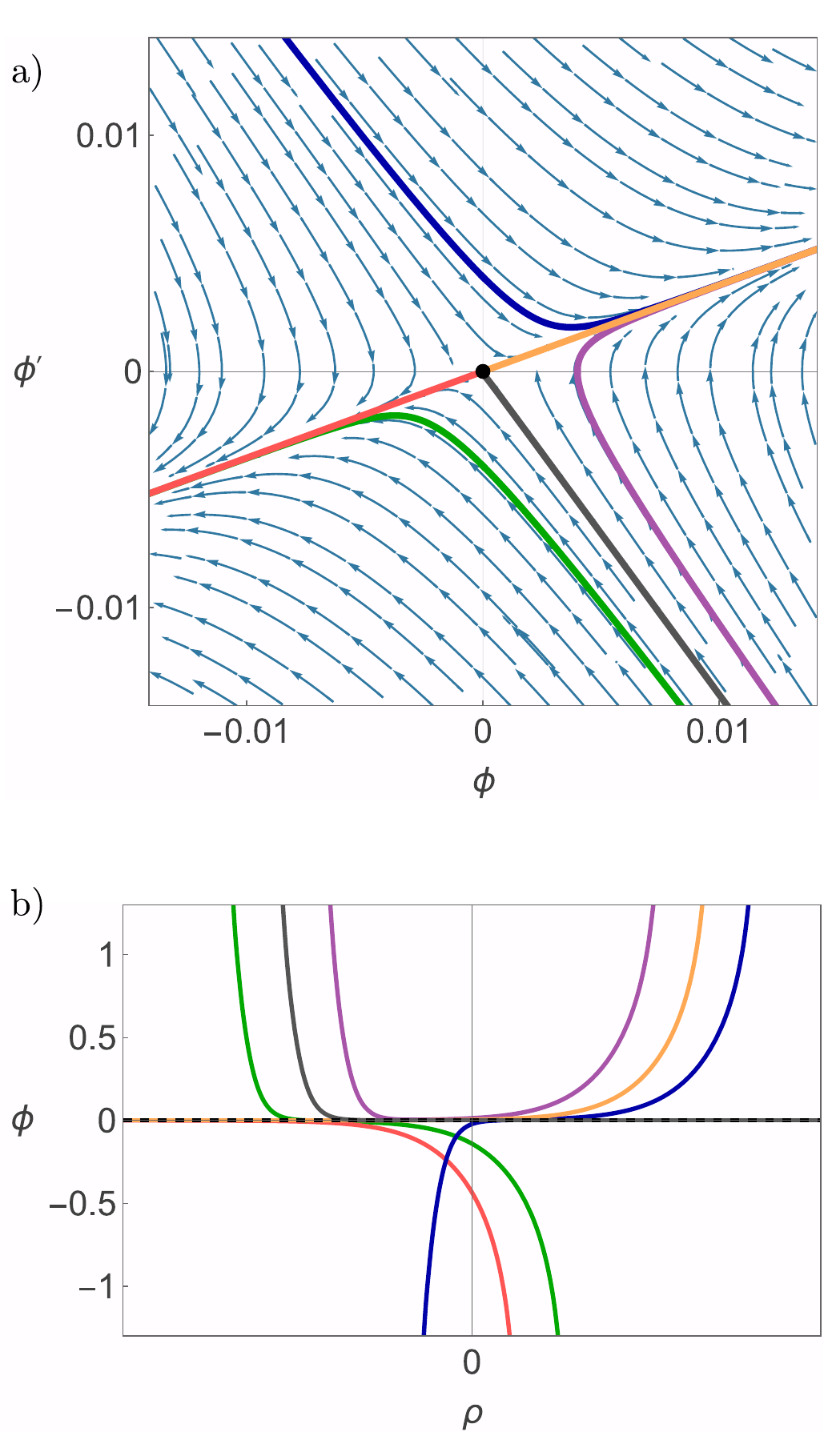}
    \caption{An illustration of the (a) phase trajectories and (b) spatial profiles of the order parameter in the Poincar\'e half-plane for $\tau<0$ (here we set $R=1$ and $\tau = -1/2$), i.e., no superconductivity in the~bulk.} 
    \label{fig:GLPhase3}
\end{figure}

In the case of a finite-$\rho_0$ boundary, we again find three options. 
First, we can have either a segment of the constant trajectory (if $\phi(\rho_0) = \phi_b$, dashed black in Fig.~\ref{fig:GLPhase2}(b)) or of the oscillating trajectory (if $\phi(\rho_0) < \phi_b$, red in Fig.~\ref{fig:GLPhase2}). 
In contrast to the case $4R^2\,\tau<1$ considered before, the solution here 
also happens to be possible with the boundary condition $\phi_b=0$.) 
In addition, as before, we can hit another attractive direction of the saddle and find a monotonic profile (if $\phi(\rho_0)>\phi_b$, orange line in Fig.~\ref{fig:GLPhase2}). Therefore, for both cases of $\tau$, the GL approach allows for both a boundary-suppressed and a bounary-enhanced superconductivity.

\subsubsection{Poincaré half-plane for $\tau<0$} 

We conclude the analysis of GL equations on the Poincaré half-plane by giving a closer look at the case of $\tau<0$, which in the absence of boundaries corresponds to the normal conductor phase. Indeed, the system \eqref{eq:dynamical_system} admits a single fixed point $(q_0,\,p_0) = (0,\,0)$. The exponents are real and of opposite sign,
\begin{equation}
\lambda_{1,2}=\frac{1}{2R}\left(-1\pm\sqrt{1+4R^2\tau}\right),
\end{equation}
implying that this is a saddle point.

In the case of the entire plane, there is a single admissible profile for the order parameter: the trivial profile $\phi=0$. 
However, more possibilities can be found when boundaries are introduced. In the case of a finite cutoff near the external boundary, one can also sit on one of the attracting separatrixes of the saddle (red and orange trajectories in Fig.~\ref{fig:GLPhase3}). In this case, the order parameter is maximized at the boundary, and exponentially suppressed in the bulk: this regime can be interpreted as the boundary-only superconductivity, and may be compared with the results reported in Fig.~\ref{fig:Delta_Caylee_tree_effective}. 

\subsection{Annulus and disc geometries}
\label{sec:GL-annulus}

Having discussed the auxiliary case of the half-space, we now turn to the actual geometry of interest and start with the annulus. Here, we rely on the fact that, in the limit $r\gg R$, the Poincar\'e disc metric \eqref{Disc_Metric} turns into the half-plane metric \eqref{HalfPlaneMetric}; therefore, for an annulus with both boundaries being of sufficiently large radius (which is what we consider in what follows), we can refer to the results obtained in the half-plane coordinates, despite the non-precise equivalence between two theories.
In practice, this appears to be a reasonable approximation already for $r/R\leq 1$. Note that the disc of radius $r/R<1$ (where our approximation 
breaks down) is rather small in the sense that for the $\{7,3\}$ and $\{8,3\}$ lattices it contains a single elementary cell. For other hyperbolic lattices, this disk
is instead contained in a single elementary cell. Such approximation of the metric should therefore be sufficient for comparison with the physics of hyperbolic lattices containing multiple polygons. For this reason, we will adopt this approximation in what follows.

The fact that both infinities $\rho\rightarrow\pm\infty$ (we are now using again the half-plane coordinate $\rho$) are cut off relaxes the regularity conditions, and more types of trajectories are allowed. For instance, there is an option when the order parameter takes finite values of the same sign at the boundaries, while being suppressed in the bulk (the purple trajectory in Figs.~\ref{fig:GLPhase1}-\ref{fig:GLPhase3} for same sign at both boundaries, and blue trajectory in Figs.~\ref{fig:GLPhase1}-\ref{fig:GLPhase3} together with the green trajectory in Fig.~\ref{fig:GLPhase3} for opposite signs at the boundaries). We can also find a situation where the order parameter takes finite values of the opposite signs at the boundaries, while crossing the zero value at some points in the bulk (green and blue trajectories in Fig.~\ref{fig:GLPhase3}). 
In addition, the case of boundary-suppressed superconductivity at both edges of the annulus is also possible (the green curves in Figs.~\ref{fig:GLPhase1}-\ref{fig:GLPhase2}, assuming an appropriately chosen range of $\rho$ where $\abs{\phi} < \sqrt{2\tau}$).

Recapitulating the above discussion, we can also explain our claim of why the annulus is a more suitable geometry for comparison with the lattice. The main difference brought up by the annulus geometry is that the bulk regularity condition is released. On the other hand, the same is expected from the lattice, where the discrete structure mitigates the possible divergences. In this regard, the inner boundary of the annulus plays conceptually the same role as the UV cutoff imposed by the lattice constant in the discrete hyperbolic structures. 

Finally, for the sake of completeness, we also consider radially symmetric solutions to the GL functional on the Poincar\'e disc: considering finite disks as well as the entire infinite plane 
with a conformal boundary. 
We focus in the remainder of the section on the bulk-superconducting case with $\tau>0$.
This geometry happens to be more difficult for a direct analysis, since one would have to deal with a non-linear, non-autonomous equation, we therefore do not have an ambition of providing the complete classification of the possible types of behaviour of the order parameter. Nevertheless, we will argue that certain familiar types of behaviour  can also be recovered here. Our strategy will be to approximate the solutions by a  solution of a linearized equation for $r\ll r_*$, and by a solution with the approximate metric Eq. \eqref{HalfPlaneMetric} for $r>r_*$ for a certain stitching point $r_*$.

The preceding analysis does not provide the correct picture when we are close to the center of the disc, $r\ll R$. In this case, the metric can be approximated by
\begin{equation}\label{Flat_Disc_Metric}
    g_{ab}=\left( 
    \begin{matrix}
        1 && 0\\
        0 && r^2
    \end{matrix}
    \right),
\end{equation}
which is just the flat-space metric. Analysing the behaviour of the order parameter in this region, we will also assume that it is close to its ground-state value, $\phi=\sqrt{2\tau}+\xi$, $\xi\ll\sqrt{2\tau}$, such that the equation of motion can be linearized:
\begin{equation}
    \xi''+\frac{1}{r}\xi'-2\tau\xi=0.
\end{equation}
Imposing regularity at the origin, we find the solution
\begin{equation}
    \xi=C\,\mathcal{I}_0(\sqrt{2\tau}\,r) \quad \textrm{for $r\ll R$},
\end{equation}
where $\mathcal{I}_n(x)$ is the modified Bessel function of the first kind, and $C$ is a constant determined by boundary conditions. Note that the value of the order parameter at the origin $\phi(r=0)=\sqrt{2\tau}+C$ never reaches the ground-state value (unless we are actually dealing with the homogeneous solution, where $\xi\equiv0$), but it may be smaller or greater than $\phi_b$, depending on the sign of $C$.

\begin{figure}[t]   
    \includegraphics[width=0.42\textwidth]{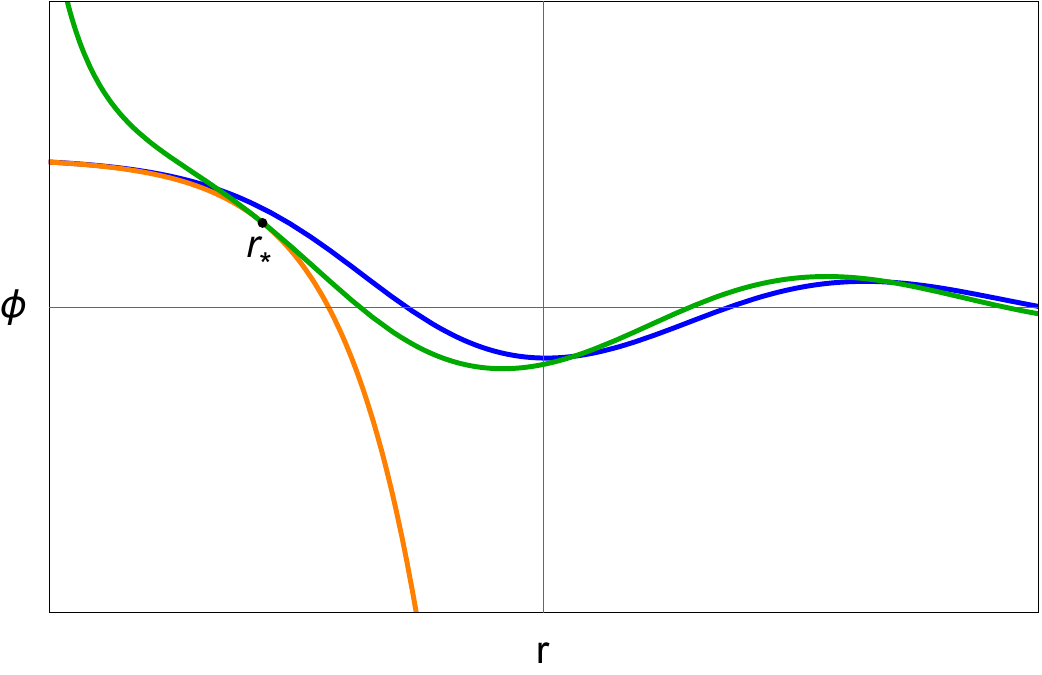}
    \caption{An illustration of how the exact solution of the Eq.~\eqref{AdS_Scalar_EOM} with the disc metric Eq.~\eqref{Disc_Metric} (blue curve) can be approximated by a solution of the linearized equation Eq.~\eqref{eq:Disc_Linearized} (orange curve) for $r<r_*$, and by a solution with the approximated metric Eq.~\eqref{HalfPlaneMetric} (green curve) for $r> r_*$, where $r_*$ is the stitching point. 
    The curves are plotted for $\tau R^2=6$ and $C = -0.02$.}
    \label{fig:GLApprox}
\end{figure}

In fact, the linearized equation can be solved with the full Poincar\'e metric. The equation takes the form
\begin{equation}\label{eq:Disc_Linearized}
    \xi''+\frac{\coth \tfrac{r}{R}}{R}\,\xi'-2\tau\,\xi=0,
\end{equation}
and the solution, regular at the origin, is given in terms of the Legendre functions of the first kind:
\begin{equation}
\label{eqn:GL-disk-solution}
    \xi=C\,P_{\nu}(\cosh\tfrac{r}{R}),\qquad \nu=\frac{-1+\sqrt{1+8\tau R^2}}{2},
\end{equation}
where $C$ is again an arbitrary constant.\footnote{When $\nu\in\mathbb{Z}$, or equivalently $\tau R^2 = \tfrac{n(n+1)}{2}$, the Legendre function happens to be polynomial (the Legendre polynomial).} Now, for any finite radius 
$r_*$ we can choose $C$ to be small enough, such that the linear approximation is valid up to $r_*$. If in turn $r_*$ is sufficiently large, we can use the approximate metric \eqref{HalfPlaneMetric} to follow the solution for 
$r>r_*$, and consequently the discussion for the half-plane is applicable (see Fig.~\ref{fig:GLApprox} for an illustration of this approximation scheme). It is a simple matter to check that when the constant $C$ is negative and small enough, the point on the phase space $(\phi(r_*),\,\phi'(r_*))$, from which the non-linear stage is supposed to start, will lie in the basin of attraction of the stable fixed point at the origin. In the case of $4R^2\tau<1$, and again assuming that $C$ is sufficiently small, the order parameter is expected to decrease monotonically toward the boundary, similarly to the red trajectory or the lower branch of the green trajectory of Fig.~\ref{fig:GLPhase1}  (recall that from $r=r_*$ onward we can use the half-plane picture). On the other hand, for larger in absolute value (but still negative) $C$ we may expect trajectories with an oscillation. 
When $C$ is positive, we find a trajectory corresponding to the order parameter increasing monotonically toward the boundary (similarly to the orange or purple lines in Fig.~\ref{fig:GLPhase1}). Similar considerations apply to the phase with $4R^2\tau>1$, with the main difference that the monotonic trajectories for $C<0$ are replaced by the oscillating~ones.

\section{Conclusion}
In this paper, we have studied superconductivity on hyperbolic lattices using complementary approaches that span from discrete lattice models to continuum field theory. Our investigation reveals that hyperbolic geometry fundamentally changes the physics of superconducting systems, particularly through boundary effects that have no analog in flat space.
Our analysis of finite hyperbolic lattices using Cayley-tree approximations shows that the radial symmetry inherent to tree structures provides a useful framework for understanding boundary superconductivity. We found that small hyperbolic flakes exhibit order parameter profiles remarkably similar to those of Cayley trees with appropriate effective connectivity. The self-consistent equations derived in this approximation successfully capture the enhancement of the superconducting gap near boundaries, which emerges because hyperbolic lattices maintain a finite fraction of boundary sites regardless of system size. However, this approach has clear limitations: it works well for even $p$-lattices around half-filling but breaks down for odd $p$-lattices or when the chemical potential deviates significantly from zero, revealing the importance of loop structures that are absent in pure tree geometries.

By numerically solving the BdG equations, we scanned through the full parameter space to reveal how superconductivity depends on the geometric properties of hyperbolic lattices. First, by considering diverse $\{p,q\}$, we found that the superconducting phase diagrams closely follow the corresponding density of states, even in these finite inhomogeneous systems where boundary effects are expected to dominate. This suggests that the boundary contributions effectively determine the density of states in finite hyperbolic flakes. We also discovered that the method of constructing finite flakes significantly affects the results: site-centered flakes with dangling boundary sites exhibit dramatically enhanced superconductivity near $\mu=0$ due to zero-energy modes associated with dangling site configurations. This enhancement disappears when the dangling sites are removed, highlighting the sensitivity of superconducting properties to microscopic boundary details.

The continuum Ginzburg-Landau analysis on the hyperbolic plane provides a complementary perspective that goes beyond lattice-specific details. A key insight from this approach is that finite geometries with boundaries (such as discs or annuli) enable radial variations of the superconducting condensate that are impossible in infinite space. By studying the phase portraits of the order parameter dynamics, we identified distinct regimes depending on the temperature parameter $\tau$. For $4R^2\tau < 1$, the system exhibits monotonic profiles, while for $4R^2\tau > 1$, oscillatory behavior emerges. Crucially, the presence of boundaries allows for boundary conditions that can drive strong enhancement of the order parameter near the edges. Both temperature regimes support solutions where the superconducting order parameter increases toward the boundary, demonstrating that this enhancement is enabled specifically by the boundary conditions rather than being a bulk property. This agreement between the lattice and the continuum approaches confirms that boundary enhancement is an intrinsic feature of hyperbolic geometry that emerges when appropriate boundary conditions are imposed on finite systems.

Several questions remain open for future work. The interplay between hyperbolic geometry and other forms of superconducting order beyond 
$s$-wave pairing is largely unexplored. In particular, $d$-wave superconductivity, with its nodal structure and directional dependence, could exhibit fundamentally different behavior on hyperbolic lattices where the local coordination environment varies systematically with distance from boundaries. Beyond the mean-field treatment employed here, quantum phase fluctuations of the superconducting order parameter represent another important frontier. The enhanced boundary density of states in hyperbolic systems could significantly modify the strength and character of these fluctuations compared to flat space, potentially leading to novel quantum critical phenomena. Similarly, going beyond mean-field theory using approaches such as real-space GW methods could reveal correlation effects that are masked in the BdG approximation, particularly near boundaries where the local environment deviates most strongly from bulk behavior.

The repulsive Hubbard model on hyperbolic lattices opens up a different set of questions. Besides superconductivity, it can give rise to magnetic phases with unique characteristics imposed by the hyperbolic geometry. Recent work has shown that magnetism in hyperbolic lattices can exhibit quite peculiar features~\cite{Ortiz:2025}, suggesting that boundary effects in hyperbolic spaces could lead to magnetic orders with no flat-space analogs. The competition between different magnetic orderings and the role of geometric frustration in hyperbolic tilings remain largely unexplored.
Additionally, the role of disorder and its interaction with the enhanced boundary density of states in hyperbolic systems presents interesting possibilities for engineering superconducting properties. The combination of hyperbolic geometry with topological considerations could also open new directions for realizing exotic superconducting phases in curved spaces.

\section{Acknowledgments}

We thank M.~Katsnelson, M.~Pavliuk, and M.~Titov for valuable discussions.
A.I. acknowledges support from the UZH Postdoc Grant, grant No. FK-24-104.
T.B., and A.I~were supported by the Starting Grant No.~211310 by the Swiss National Science Foundation. A.A.B. was supported by by NWO grant
OCENW.M.23.044 and the research program ``Materials for the Quantum Age'' (QuMat). This
program (registration number 024.005.006) is part of the
Gravitation program financed by the Dutch Ministry of
Education, Culture and Science (OCW).

\appendix
\section{BdG equations on Cayley trees}
\label{app:Cayley_BdG}

In this Appendix, we focus on the description of (non)symmetric states and the application of symmetry-adapted basis to solutions of BdG gap-equations. The approach described here follows the discussion in Ref.~\cite{Pavliuk:2025} and is adapted in the present work for the calculations on Cayley-tree approximation of hyperbolic lattices in Section~\ref{sec:Cayley_tree_apx}, where large sizes of the tree approximants are necessary to estimate the boundary effects in the superconducting order parameter. In Sec.~\ref {app:Cayley_states}, we begin with the construction of (non)symmetric states. In Sec.~\ref {app:Cayley_BdG1}, we continue with the description of symmetry-adapted sectors of the single particle Hamiltonian. Finally, we derive the gap equation written in the basis of (non)symmetric states in Sec.~\ref {app:Cayley_BdG2}.

\subsection{Eigenstates of Caylee trees}\label{app:Cayley_states}

In the description of symmetric and nonsymmetric states, we follow Refs.~\citenum{Hamanaka:2024} and~\citenum{Pavliuk:2025}.
We consider the Cayley tree with connectivity $K$ and the total number of layers $M$.
Since the central node has $K+1$ branches and the $l$-th $(l=1, \cdots, M)$ generation of each branch has $K^{l-1}$ nodes, the total number of sites and the
dimension $\mathcal N$ of the Hilbert space are
\begin{align}
    \mathcal{N} =  1 + (K+1) \times \sum_{l=1}^M K^{l-1} = 1 + (K+1) \frac{K^M -1}{K-1}.
\end{align}
The original Hilbert space consists of position basis vectors $|i\rangle$, where the index $i$ denotes a site on a Cayley tree. We build linear combinations of these position basis states, allowing us to block-diagonalize the Hamiltonian.

\begin{figure}[t]
    \centering
    \includegraphics[width=\linewidth]{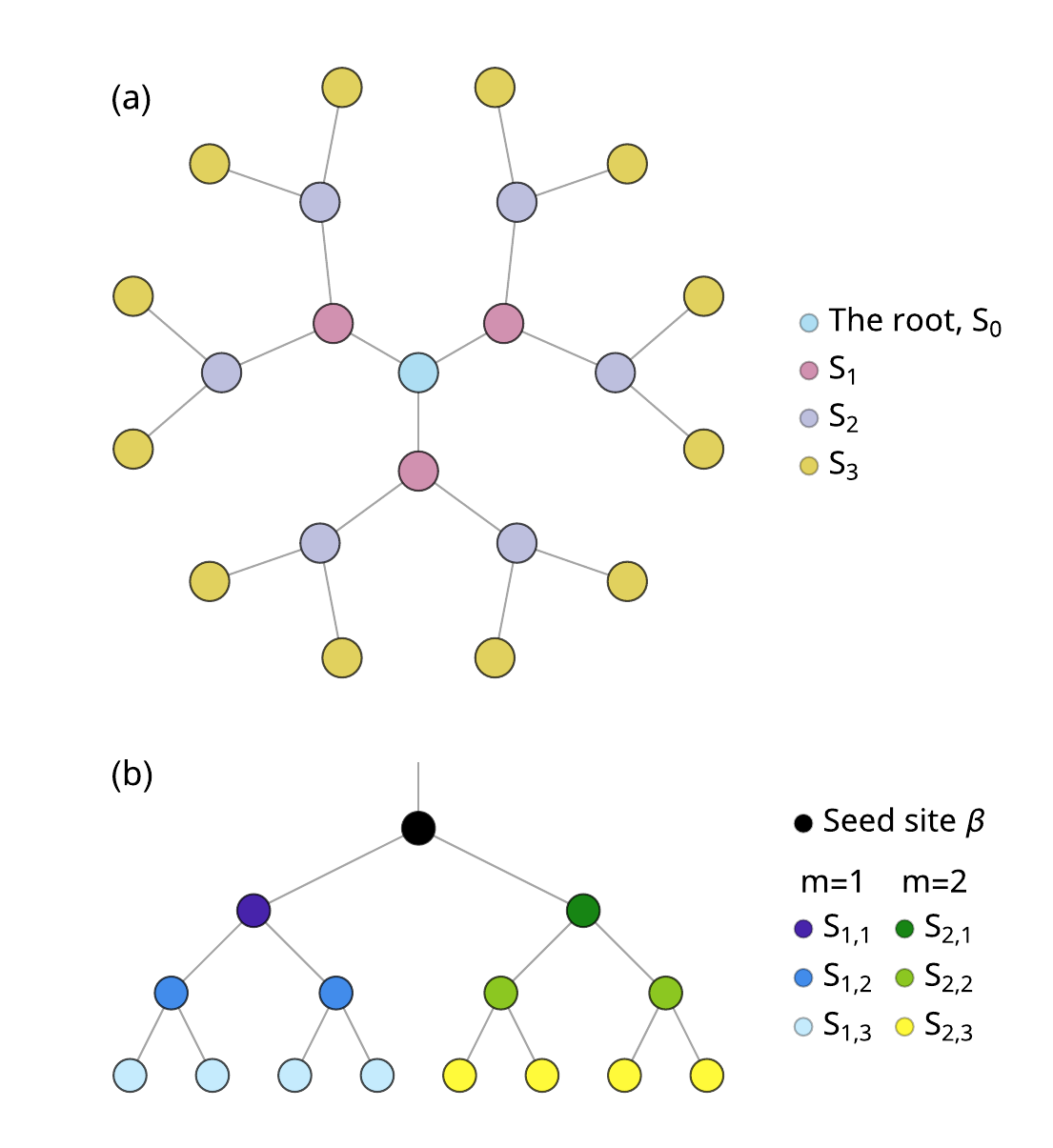} 
    \caption{Panels 
    (a) and (b) illustrate the definitions of shells for symmetric and nonsymmetric states. In panel
    (a), different shells $S_l$ are shown by different colors. The symmetric linear combination of the sites in each of the shells gives symmetric states in Eq.~\eqref{eq: symm basis}. In panel
    (b), different shells $S_{m,r}$ are shown by different colors. Since the connectivity $K=2$, the index $m$ takes values in $\{1,2\}$. To construct a non-symmetric state, we choose all sites at a distance $r$ from the `seed' node. Then we make a linear combination of these sites according to \cref{eq:nonsymstate_betaneq0,eq: nonsymm basis0,eq: nonsymm basis}. We choose the same weight for all sites with the same $m$ (the same color in the panel) in such a way that the sum of the weights is $0$.}
    \label{fig:cayley_states}
\end{figure}

We start by building the set of symmetric basis states. The first symmetric basis state is equal to the position basis state on the central (`root') site
\begin{align}
    |0) \coloneqq \ket{0}.
\end{align}
In the following discussion, by $\ket{\cdots}$ and $|\cdots)$, we denote the position basis and the symmetry-adapted (i.e., symmetric and nonsymmetric)
basis states. For the construction of the next symmetric basis states, we need to consider the generations of sites starting from the root of the tree, where we call the $l$-th generation as the $l$-th shell of the tree and label it as $S_l$. We generate the remaining symmetric basis states $| l)$ by symmetrizing the position basis states for each shell $S_l$ of the Cayley tree:
\begin{align}
\label{eq: symm basis}
    | l) \coloneqq \frac{1}{\sqrt{(K+1)K^{l-1}}} \sum_{\alpha\in S_l} |\alpha\rangle,
\end{align}
which form $M+1$ symmetric orthonormal states. The shells $S_{l}$, over which the linear combinations of position states are taken, are illustrated in Fig.~\ref{fig:cayley_states}~(a).

Next, we generate the remaining basis states, which we call nonsymmetric basis states.
We choose a node $\beta$ lying not in the last ($M$-th) shell as the `seed' and consider the $K$ branches (or $K+1$ if $\beta$ is a central node) rooted at this node. The key idea in constructing the nonsymmetric states is to weight the branches emitted from the node $\beta$ in such a way that the action of the Hamiltonian creates destructive interference of the hopping processes from the branches at the seed node. 

To be more precise, we denote the sites that are direct descendants of the site $\beta$ as $\gamma_m$, where $m$ takes values in $\{1,\ldots, K\}$, since $K$ is the number of branches starting from $\beta$, if $\beta$ is not a central site. In case if $\beta$ is a central site, $m$ takes values in $\{1,\ldots, K+1\}$, since the central site has $K+1$ branches. Then for each $|\gamma_m\rangle$, we introduce weights $c_m$, such that $\sum_m c_m=0$. The space of the first nonsymmetric states associated with the site $\beta$ is given by the linear combinations of $|\gamma_m\rangle$ with weights $c_m$
\be
|1)_\beta=\sum_m c_m|\gamma_m\rangle.
\ee
The space of these vectors is usually more than one-dimensional, so we need to introduce a basis in it. One way to do it is to choose a $K$-th root of unity for any seed $\beta$ except when the seed is the central site, and $(K+1)$-th root when $\beta$ is the central site. In both cases, we denote the root of unity as $\omega$,and we implicitly assume that possible choices of $\omega$ depend on $\beta$. Then we can take the weights $c_m$ as $c_m=\omega^m$. Having done that, we obtain the basis of the first nonsymmetric states associated with the site $\beta$

\begin{equation}
\label{eq:nonsymstate_betaneq0}
\!|1,\varpi)_\beta \!=\! 
\left\{\begin{array}{l}
\!\!\frac{1}{\sqrt{K+1}}\!\sum^{K+1}_{m=1}\! \omega^m|\gamma_m\rangle \; \textrm{with $\omega = e^{\frac{2\pi i \ell}{K+1}}$} \, \textrm{for $\beta = 0$\!} \\ 
\!\!\frac{1}{\sqrt{K}}\!\sum^{K}_{m=1}\! \omega^m|\gamma_m\rangle \;\textrm{with $\omega = e^{\frac{2\pi i \ell}{K}}$} 
\, \textrm{for $\beta \neq 0$,}
\end{array}\right.
\end{equation}
where $\ell$ is a positive integer, and we also normalize the states.
The next nonsymmetric states are obtained by considering the next generations of descendants of the site $\beta$, similarly to the symmetric states.

Finally, we can write the nonsymmetric states corresponding to the central node as
\begin{align}\label{eq: nonsymm basis0}
    |r,\omega)_0 \coloneqq \frac{1}{\sqrt{(K+1)K^{r-1}}} \sum^{K+1}_{m=1}\omega^m\sum_{\alpha\in S_{m,r}}|\alpha\rangle,
\end{align}
where by $S_{m,r}$ we denote the $r$-th shell of the branch $m$ starting from the node $\beta$, which in this case is the central site, and the multiplication prefactor is the consequence of the normalization. In this case the index $r$ takes the values in $\{1,\ldots,M\}$. For other nodes, we have accordingly
\begin{align}\label{eq: nonsymm basis}
    |r,\omega)_{\beta\neq 0} \coloneqq \frac{1}{\sqrt{K^{r}}} \sum^{K}_{m=1}\omega^m\sum_{\alpha\in S_{m,r}}|\alpha\rangle
\end{align}
where index $r$ takes the values in $\{1,\ldots,M-l_\beta\}$ and $l_\beta$ is the radial distance of $\beta$ from the center. The shells $S_{m,r}$, over which the linear combinations of position states are taken, are illustrated in Fig.~\ref{fig:cayley_states}(b).

We can explicitly check that the constructed set of symmetric and nonsymmetric states is complete and
orthonormal.
Therefore any state $\ket{\Psi}$ is expanded using these basis states as
\begin{align}
    \ket{\Psi} =\sum \psi_l |l)+\sum_{\beta\in \mathcal{T}}\sum_{r}\sum_\omega\psi_{\beta,r,\omega}|r,\omega)_\beta
\end{align}
where $\psi_0$ and $\psi_{\beta,r,\omega}$ are the wave function components, $\mathcal{T}$ denotes the whole tree except the last shell, we assume the dependence of the range of $\omega$ on the root node $\beta$, and the sum over $r$ depends on the node $\beta$ as well and is performed for all shells between the root node and the corresponding offspring boundary nodes. Finally, in this paper, we refer to eigenstates that can be expanded solely using symmetric (nonsymmetric) basis states as symmetric (nonsymmetric) eigenstates.

\subsection{Sectors of (non)symmetric states}
\label{app:Cayley_BdG1}

In this section, we describe the single particle Hamiltonians of (non)symmetric sectors to prepare the setup for discussing the gap equations in the symmetry-adapted basis.

The first thing that one can notice is that the kinetic part of the BdG Hamiltonian $h$ in Eq.~(\ref{eq:BdG_Ham}) defined on a Cayley tree can be decomposed into a block diagonal form. One block acts on symmetric states, and all other blocks act on the nonsymmetric states. We will call these blocks the symmetric and nonsymmetric blocks accordingly. Nonsymmetric states with the same seed node $\beta$ and the same root of unity $\omega$ constitute a single
block. Summing up, the Hamiltonian $h$ can be written as follows

\begin{gather}\label{eq:Ham_block_Cayley}
h = \begin{pmatrix}
h_\textrm{sym} & 0 & 0 & 0 & \cdots \\
0 & h^{0}_\textrm{non-sym} & 0 & 0 & \cdots \\
0 & 0 & h^{1}_\textrm{non-sym} & 0 & \cdots \\
0 & 0 & 0 & h^{2}_\textrm{non-sym} & \cdots \\
\vdots & \vdots & \vdots & \vdots & \ddots \\
\end{pmatrix}   
\end{gather}
where the upper index in the nonsymmetric Hamiltonian blocks $h_\textrm{non-sym}$ denotes a pair of the seed node and the root of unity $(\beta,\omega)$.

After applying the Hamiltonian $h$ to the set of symmetric and nonsymmetric states, one can find the form of the blocks $h_\textrm{sym}$ and $h_\textrm{non-sym}$. Let us begin with the symmetric states.
One can find that in this case the following relations hold
\begin{gather}
h|0)=\sqrt{K+1}|1)\\\nonumber
h|M-1)=\sqrt{K}|M)\\\nonumber
h|l)=\sqrt{K}|l-1)+\sqrt{K}|l+1), \,\,\, 0<l<M
\end{gather}
These relations hold because, by the action of the Hamiltonian on symmetric states, one obtains symmetric states on the neighboring shells. Therefore, one can write the matrix form of the symmetric block $h_\textrm{sym}$ as follows
\begin{gather}\label{eq:H_sym}
    h_\textrm{sym} = \begin{pmatrix}
0 & \sqrt{K+1} & 0 & 0 & \cdots \\
\sqrt{K+1} & 0 & \sqrt{K} & 0 & \cdots \\
0 & \sqrt{K} & 0 & \sqrt{K} & \cdots \\
0 & 0 & \sqrt{K} & 0 & \cdots\\
\vdots & \vdots & \vdots & \vdots & \ddots \\
\end{pmatrix}.
\end{gather}

In the case of non-symmetric states, the action of the Hamiltonian is similar. One can check that the following relations hold
\begin{gather}
h|1,\omega)_\beta=\sqrt{K} h|2,\omega)_\beta\\\nonumber
h|M-l_{\beta},\omega)_\beta=\sqrt{K} h|M-l_{\beta}-1,\omega)_\beta\\\nonumber
h|r,\omega)_\beta=\sqrt{K}|r-1,\omega)_\beta+\sqrt{K}|r+1,\omega)_\beta, \,\,\, 1<r<M-l_{\beta}
\end{gather}
where  $l_\beta$ is the radial distance of $\beta$ from the center.
Thus, in the matrix form, the non-symmetric blocks $h^{k}_\textrm{non-sym}$ for the non-symmetric states with given $\beta$ and $\omega$ can be written as 
\begin{gather}\label{eq:H_nonsym}
    h^{k}_\textrm{non-sym} = \begin{pmatrix}
0 & \sqrt{K} & 0 & 0 & \cdots \\
\sqrt{K} & 0 & \sqrt{K} & 0 & \cdots \\
0 & \sqrt{K} & 0 & \sqrt{K} & \cdots \\
0 & 0 & \sqrt{K} & 0 & \cdots \\
\vdots & \vdots & \vdots & \vdots & \ddots \\
\end{pmatrix},
\end{gather}
The block diagonal form of the Hamiltonian helps us in writing the BdG equations in a more compact form.

\subsection{Modified gap equation}
\label{app:Cayley_BdG2}

In the previous section, we constructed the symmetry-adapted sectors of the single-particle Hamiltonian. In this section, we derive the modified gap equation in the symmetry-adapted basis given by Eq.~\eqref{eq:selfcons_Cayley}.

Having described the (non)symmetric sectors and the corresponding blocks of the BdG Hamiltonian, we are set to incorporate the symmetries into the gap equation. The first useful property is that we should use the radial symmetry of the Cayley tree. Hence, for the solution of BdG equations, we need to choose only one site in each shell of the tree. The second fact is that nonsymmetric states have support only on some wedge of the tree. Therefore, when we consider self-consistent equations Eq.~\eqref{eq:selfcons_delta}, we should take into account only a relatively small number of nonsymmetric states that are non-zero at the chosen node. The only states that can take non-zero values at the node $i$ are the states whose seed node $\beta$ is a parent of the given node. Hence, we can rewrite the self-consistent equations as follows (see Ref.~\citenum{Pavliuk:2025} for further details)
\begin{gather}
    \Delta_i=\frac{U}{2}\sum_{\substack{\textrm{symmetric}\\\textrm{states}}}\sum_{n}u_{n,i} v^{*}_{n,_i} \tanh(\frac{E_n}{2 T})+\\
    +\frac{U}{2}\sum_{\substack{\textrm{nonsymmetric}\\\textrm{states,}\,\, \beta\in P(i)}}\sum_{n}u_{n,i} v^{*}_{n,_i} \tanh(\frac{E_n}{2 T}), 
\end{gather}
where by $P(i)$ we denote the nodes that are parents of the site $i$. In other words, symmetric states have non-zero support everywhere, so they should be taken into account for each shell. However, the further the shell lies from the center of the tree, the fewer nonsymmetric states we need to take into account.

We can also note that there is a degeneracy of non-symmetric states coming from the different roots of unity $\omega$. The contributions to the $\Delta_i$ coming from the states with different $\omega$, but the same $\beta$ are equal, since $\omega$ only changes a phase. This means that for the solution of BdG equations, we can consider only the sector corresponding to symmetric states and $M$ sectors of nonsymmetric states, where each nonsymmetric sector $i$ corresponds to a node within the $i$-th shell. After including the degeneracy multiplier, we can write the self-consistency equations as
\begin{gather}
    \label{eq:selfcons_tree_orbasis}
    \Delta_i=\frac{U}{2}\sum_{\substack{\textrm{symmetric}\\\textrm{states}}}\sum_{n}u_{n,i} v^{*}_{n,_i} \tanh(\frac{E_n}{2 T})+\\
    +\frac{U K}{2}\sum_{\substack{\textrm{nonsymmetric}\\\textrm{states,}\,\, \beta=0}}\sum_{n}u_{n,i} v^{*}_{n,_i} \tanh(\frac{E_n}{2 T})+\\
    +\frac{U(K-1)}{2}\sum_{\substack{\textrm{nonsymmetric}\\\textrm{states,}\,\, \beta>1}}\sum_{n}u_{n,i} v^{*}_{n,_i} \tanh(\frac{E_n}{2 T}). 
\end{gather}

Summing up, the BdG equations on a Cayley tree can be solved, if one considers the kinetic Hamiltonian $h$ in block form given by Eq. \eqref{eq:Ham_block_Cayley} and takes into account only the symmetric block $h_{\textrm{sym}}$ and $M$ nonsymmetric blocks $h^{k}_{\textrm{non-sym}}$.  However, the self-consistent equations in Eq. \eqref{eq:selfcons_tree_orbasis} are written in the original position basis, while
the Hamiltonians for the symmetric and nonsymmetric blocks $h_{\textrm{sym}}$ and  $h^{k}_{\textrm{non-sym}}$ are written in the basis of symmetric and nonsymmetric states. Introducing new BdG vectors $(\widetilde{u}^{sym}, \widetilde{v}^{sym})$ for the symmetric sector and $(\widetilde{u}^k, \widetilde{v}^k)$ for $k$-th nonsymmetric sector, the self-consistent equations take the final form
\begin{gather}
\label{eq:selfcons_Cayley}
    \Delta_i=\frac{U}{2}\Big[\sum_{n; i\ge 0}N^{0}_i \widetilde u^{sym}_{n,i} \widetilde v^{sym}_{n,i} \tanh(\frac{E^0_n}{2 T})+\\\nonumber+\frac{K^2}{K+1}\sum_{n;i>0} K^{-i}\widetilde u^{0}_{n,i} \widetilde v^{0}_{n,i} \tanh(\frac{E^1_n}{2 T})+\\\nonumber+(K-1)\sum_{n; i> k> 0}K^{k-i-1}\widetilde u^{k}_{n,i} \widetilde v^{k}_{n,i} \tanh(\frac{E^k_n}{2 T})\Big],
\end{gather}
where additional weights appear in comparison with Eq. \eqref{eq:selfcons_tree_orbasis} since (non)symmetric states are obtained via averaging over the shell in the position basis. The constraints on the indices $i$ are written explicitly to outline the fact that the nonsymmetric states contribute only to the sites with $i>\beta$. We also omit the complex conjugation in the self-consistent equations, since we can gauge $\Delta_i$ to be real, and therefore take the real eigenvectors as well.
The weight for the symmetric states $N^0_i$ is defined as
\begin{gather}
    N^0_i=\begin{cases}
        1, \quad\textrm{if}\quad i=0\\
        \frac{1}{(K+1)K^{i-1}} , \quad\textrm{if}\quad i>0
    \end{cases}
\end{gather}

The presented calculation scheme allows us to calculate the self-consistent order parameter on trees with radial size up to a few hundred shells. We conclude by noticing that in the Cayley-tree approximation of hyperbolic lattices, we use \cref{eq:selfcons_Cayley} but with effective values $K_l$ that depend on the shell in a manner that mimics the growth of the hyperbolic lattice. 

\bibliography{bib}

\end{document}